\documentclass[prb,aps,floats,amssymb,showkeys,showpacs,preprint,
superscriptaddress,tightenlines]{revtex4}
\usepackage{graphicx}
\usepackage{epsfig,bm}
\begin{document}
\preprint{Bicocca-FT   June 2009}

\title{ 
Extended scaling behavior of the spatially-anisotropic classical \\
 $XY$  model in the crossover from three to two dimensions\\
}
\author{P. Butera\cite{pb}}
\affiliation
{Istituto Nazionale di Fisica Nucleare \\
Sezione di Milano-Bicocca\\
 3 Piazza della Scienza, 20126 Milano, Italy}

\author{M. Pernici\cite{mp}}
\affiliation
{Istituto Nazionale di Fisica Nucleare \\
Sezione di Milano\\
 16 Via Celoria, 20133 Milano, Italy}

\date{\today}
\begin{abstract}
The bivariate high-temperature expansion of the spin-spin
correlation-function of the three-dimensional classical $XY$ (planar
rotator) model, with spatially-anisotropic nearest-neighbor couplings,
is extended from the 10th through the 21st order. The computation is
carried out for the simple-cubic lattice, in the absence of magnetic
field, in the case in which the coupling strength along the $z$-axis
of the lattice is different from those along the $x$- and the
$y$-axes.  It is then possible to determine accurately the critical
temperature as function of the parameter $R$ which characterizes the
coupling anisotropy and to check numerically the universality, with
respect to $R$, of the critical exponents of the three-dimensional
anisotropic system.  The analysis of our data also shows that the
main predictions of the generalized scaling theory for the
crossover from the three-dimensional to the two-dimensional critical
behavior are compatible with the series extrapolations.
\end{abstract}

\maketitle

\small
\section{Introduction}
The three-dimensional($3D$) layered magnetic spin systems in which the
 strength of the interactions among the layers is much smaller than
 within the layers, are often referred to as quasi-two-dimensional.
 Although strictly two-dimensional($2D$) magnetic systems do not exist
 in nature, their statistical mechanics can be studied by
 experimenting with diverse more mundane structures and in particular
 by exploring how quasi-two-dimensional systems\cite{bramwell} 
 crossover, i.e.  change their  universality class, on going from
 the $3D$ to the $2D$ critical regime.  From a more general standpoint,
 the study of spatially anisotropic systems also provides the simplest
 example of a wide variety of crossover phenomena
\cite{fishmod,pft,riedel,fishjas,pfeuty}
 of different origin which may occur near criticality.

The simplest Hamiltonian which can model a quasi-two-dimensional
 magnetic system, in the  absence of a magnetic field, is that of 
an $N$-vector spin model with axially anisotropic couplings 
\begin{equation}
H_{an} \{ v \} = -N{J_1}  \sum_{nn(xy) } 
\vec v({\vec r}) \cdot \vec v({\vec r\;'}) -  N{J_2} 
\sum_{nn(z) } 
\vec v({\vec r}) \cdot  \vec v({\vec r\;'} )
\label{tnhamilt}
\end{equation}
  We have indicated by $\vec v({\vec r})$ an $N$-component classical
 spin-vector of unit length located at the site ${\vec r}$ of a
 simple-cubic ($sc$) lattice. The first sum in (\ref{tnhamilt}) is
 extended to the nearest-neighbor ($nn$) spin pairs within each
 horizontal {\it (xy)} layer, while the second sum is over the $nn$
 spins in adjacent layers along the {\it z}-direction.  We shall
 denote by $R=J_2/J_1$ the ratio of the interlayer to the intralayer
 coupling strength which characterizes the spatial anisotropy of the
 spin couplings and therefore is sometimes referred to as {\it
 anisotropy parameter}. The thermodynamic quantities of the model can
 be expressed as functions of the variables $K_1=J_1/kT$ and
 $K_2=J_2/kT$, with $k$ the Boltzmann constant.  One may,
 equivalently, choose either the pair of variables $K_1$ and $R$ or
 the pair $K_2$ and $\bar R=1/R$.  For $R \rightarrow 0$, the system
 becomes a stack of non-interacting spin layers.  When $R=1$, the
 system has directionally isotropic interactions.  For $R \rightarrow
 \infty$, or equivalently for $\bar R \rightarrow 0$, it reduces to an
 array of non-interacting spin chains.

Only a few pioneering 
studies\cite{oitmaa,rapaport,stanley,lambeth,liustan,citteur} 
of the Hamiltonian (\ref{tnhamilt}) by high-temperature(HT) methods
are presently available. They were aimed at:

1)  a numerical test
of the critical universality\cite{griff} for the
anisotropic system, in particular of the $R$-independence
 of the critical exponents as long as $R>0$;

2)  an investigation of the change of universality class
 of the critical transition as $R \rightarrow 0$, i.e. of the
 crossover from the $sc$ to the square-lattice critical behavior.

  They relied on HT series expansions, in terms of the two variables
 $K_1$ and $K_2$, computed through 11th order\cite{oitmaa} for $N=1$
 (the spin $1/2$ Ising model) and through 10th order\cite{lambeth} for
 $N=2$ (the planar rotator or $XY$ model) or $N=3$ (the classical
 Heisenberg model), on the $sc$ lattice. The corresponding expansions
 for the face-centered-cubic lattice also reached the same orders.
 Altogether 78 coefficients were computed in the $sc$-lattice Ising
 case and 66 coefficients in the other cases but, unfortunately, no
 higher order coefficients were added since.  In the Ising case,
 although rather short, the expansions are sufficiently well behaved
 that their extrapolations can support unambiguously the simplest
 theoretical expectations concerning the crossover.  The accuracy of
 the first series analyses in the Ising case could be only marginally
 improved\cite{stilck} by resorting to bivariate Partial-Differential
 approximants\cite{fisherkerr,Gutt} (i.e. by approximately resumming
 the HT expansions in terms of the solution of a linear first-order
 {\it partial} differential equation with appropriately chosen
 bivariate polynomial coefficients) instead of using the conventional
 ratio or Pad\'e approximant (PA) methods. The reason of this failure
 is that there is no substitute for significantly longer expansions.
 Later on, also several MonteCarlo simulations\cite{yuris,lee} were
 carried out in an attempt at further clarifying the crossover
 behavior of Ising systems, but the accuracy of the results is still
 subject to controversy for small positive $R$, i.e.  in the region of
 main interest. On the other hand, for the models with $N>1$, the
 earliest HT analyses were inconclusive, thus calling for a
 substantial extension of the expansions. In particular, the crossover
 issue remained to be studied, because even the existence of a
 critical point at nonzero temperature and the nature of the critical
 singularity in the $2D$ limit were not yet firmly assessed at the
 time of the first analyses. Also the successive simulation studies,
 carried out when the critical behaviors in $2D$ of the $N=2$
 model\cite{bkt} (and of the $N>2$ models\cite{polya}) were better
 understood, probably cannot yet be considered sufficiently
 accurate\cite{chui,kawa} or are not directly comparable\cite{janke}
 with the series analyses.

  We have been motivated by this situation to devote our study to the
 $N=2$ anisotropic model, taking advantage of our recent extensions
 through order 21, (i.e. from 66 to 253 series coefficients), of the
 bivariate HT expansions for the two-spin correlation-function and its
 moments on the $sc$ lattice.  Another reason of interest into this
 model is that it has been suggested to provide an approximate description of
 layered high-$T_c$ superconductors\cite{htcs}.

 The paper is organized as follows.  In Sec.II, we briefly mention the
algorithm adopted and specify the results of our series computations.
In Sec.III, we discuss some of the simplest predictions of the
extended phenomenological scaling theory for the crossover from the
$3D$ to the $2D$ (and from the $3D$ to the $1D$) critical behavior.
We begin by reviewing in some detail the well studied $N=1$ case only
to recall the general ideas of this approach and to contrast its
features with those of the less studied and more complex $N \geq 2$
cases.  In Sec.IV, we outline our numerical analysis of the expansions
and compare the results with the theoretical expectations.

 We should finally notice that, throughout this report, for reasons of
 clarity, we have used a notation sometimes different and heavier, but
 more detailed and perhaps more explicit, than that generally adopted
 in the earliest studies of the crossover phenomena.

\section{ Extended high-temperature expansions}

 Our computation of the series coefficients was carried out using a
 computerized recursive algorithm based on the Schwinger-Dyson
 equations\cite{march}. This method was initially applied only to the
 determination of single-variable HT expansions. Only recently, taking
 advantage of the great improvements of the computer performances of
 the last decades, it could be straightforwardly
 adapted\cite{bplif,bpunp} to derive also the more memory-demanding
 and computationally-intensive bivariate expansions for a wide class
 of isotropic and anisotropic $XY$ models with $nn$ and next-to-$nn$
 interactions. It should be noted that in our approach only
 extended-integer exact arithmetic is used, thus avoiding all roundoff
 errors which limited the precision of the preceding\cite{lambeth}
 floating-point computations. To give an idea of the performance of
 the algorithm, let us note that an ordinary single-processor desktop
 personal-computer({\it pc}) can complete, in less than a second, all
 10th-order calculations for the anisotropic $N=2$ case so far
 documented\cite{lambeth} in the literature. The calculation of the
 next nine orders takes a few days. To compute the last two orders, we
 have used a {\it pc}-cluster for a time equivalent to approximately
 six months of a single {\it pc}.

Fixing $N=2$, we have calculated the spin-spin correlations
\begin{equation}
C(\vec 0, \vec x; N;K_1,R)= <\vec v(\vec 0) \cdot \vec v(\vec x)>,
\label{corf}
\end{equation}  
 for all values of $\vec x$ for which the HT expansion coefficients
are non-trivial within the maximum order reached.  As usual, here
$<O>=Tr{(O exp[-H_{an}])}/Tr{(exp[-H_{an}])}$.

In terms of (\ref{corf}), we have formed the expansions of the $l$th
order spherical moments of the correlation-function:
\begin{equation}
{ m^{(l)}(N;K_1,R)} = \sum_{\vec x  }|\vec x|^l 
< \vec v(\vec 0) \cdot  \vec v(\vec x)>
\label{sfermoms}
\end{equation}
\noindent and, in particular, of the  reduced ferromagnetic 
susceptibility defined as $\chi(N;K_1,R)={ m^{(0)}}(N;K_1,R)$.

The  second-moment correlation-length is expressed, 
 in terms of $ {m^{(2)}}(N;K_1,R)$ and $\chi(N;K_1,R )$ as
\begin{equation}
[\xi(N;K_1,R)]^2 ={ m^{(2)}}(N;K_1,R)/2d\chi(N;K_1,R).
\label{corlen}
\end{equation}
where $d$ is the lattice dimensionality i.e. $d=1$ for $\bar R=0$, 
$d=2$ for $R=0$ and $d=3$ otherwise.

 Usually, for the univariate HT expansions, the only accessible
 validation procedure of an extended computation is the comparison
 with the lower-order results that might be already known.  In the
 case at hand, only the expansion coefficients of the second moment of
 the correlation-function are tabulated through 10th order in
 Ref.[\onlinecite{lambeth}], but they contain small roundoff errors in
 the 8th figure at highest order. After correction of these errors
 they agree with our results.  Of course, this is not a very stringent
 test of correctness.  However, the extended expansions can be
 subjected, at all orders, to additional tests, some of which deriving
 from equations of Section III.  First, we can check that, taking
 $R=1$, the coefficients of the single-variable expansions for the
 corresponding quantities of the isotropic $3D$ $XY$ model, already
 known\cite{bcoen} through order 21, are reproduced. Moreover, we can
 observe that for $R=0$ and $\bar R=0$, the expansions of
 $\chi(N;K_1,R)$ and $m^{(2)}(N;K_1,R)$ reduce, as they should, to
 those of the corresponding quantities in the $2D$\cite{bperxy,aris}
 and the $1D$ $XY$ model, respectively. Finally, we take advantage of
 eqs. (\ref{parder}) and (\ref{parder1d}) of the next Section, in the
 case of the susceptibility, (or eqs. (\ref{parderm}) and
 (\ref{parderm1d}) for the second moment), to pin down two more among
 the $r+1$ series coefficients occurring at the $r$th order. The
 success of this variety of tests, through all orders we have
 computed, strengthens the confidence that our extensions of the
 bivariate expansions are correct.

Our series data for the $nn$ correlations, the susceptibility and the 
second moment of the correlation function are tabulated in an appendix,
 which for editorial reasons was not included in the printed version
\cite{bpercross} of this paper.

\section{Dimensional crossover}

For all values of $N$, at fixed $R>0$, the $3D$ spin models described
 by the Hamiltonian eq.(\ref{tnhamilt}) display a conventional
 power-law critical transition. As the reduced deviation
 $\tau(N;R)=1-K_1/K_{1c}(N;R) $ from the critical point $K_{1c}(N;R)$
 of the  $3D$ system with anisotropy parameter $R$ 
 tends to zero from above, for the
 susceptibility one has $\chi(N;K_1,R) \sim
 [\tau(N;R)]^{-\gamma(N;R)}$, while for the correlation-length one has
 $\xi(N;K_1,R) \sim [\tau(N;R)]^{-\nu(N;R)}$.  The universality
 hypothesis\cite{griff} dictates that, for a given value of $N$, the
 critical exponents $\gamma(N;R)$ of the susceptibility and $\nu(N;R)$
 of the correlation-length of the $N$-vector system with arbitrary
 finite anisotropy should be independent of $R$ as long as $R>0$ and
 thus should coincide with the exponents $\gamma(N;1)$ and $\nu(N;1)$
 of the isotropic system.  The initial part of our analysis of the HT
 expansions in the subsections A and B of Section IV, will be devoted
 to the determination of the critical temperature and exponents as
 functions of $R$ for $R>0$, thus making it possible to test
 numerically the universality of the exponents with respect to $R$.

 When $R \rightarrow 0$ the $sc$ lattice system crosses over to a
 stack of uncoupled square-lattice systems and we expect that an
 anomalous behavior at criticality indicates a discontinuous change of
 universality class.  The crossover behavior is described by a
 phenomenological scaling theory introduced in
 Ref.[\onlinecite{riedel}] and subsequently extended and clarified in
 Refs.[\onlinecite{fishjas,fishmod,pfeuty}].  This approach, whose
 validity has been verified in the mean-field approximation and in the
 spherical model\cite{fishjas,pfeuty}, is first outlined for $N=1$
 (the Ising model) in the subsection A and then generalized to cover
 also the case $N \geq 2$, in the subsection B.  The predictions of
 the scaling theory for $N=2$ will finally be compared to the HT-based
 approximations in the subsection C of Section IV.

   An anomalous behavior is also expected to occur in the 
$\bar R \rightarrow 0$ limit in which the $sc$ lattice crosses over 
to the linear lattice.

\subsection{The $N=1$ model}

For $N=1$, both the $3D$ and the $2D$ $N$-vector models display a
power-law critical behavior. As a consequence, for all critical
exponents $\gamma(1;R)$, $\nu(1;R), \ldots $, of the $3D$ Ising model
with anisotropy $R$, the $R \rightarrow 0$ limit exists and yields the
corresponding exponents $\gamma(1;0)$, $\nu(1;0), \ldots $ of the $2D$
Ising model.  Thus, for $R = 0$, one can write
\begin{equation}
\chi(1;K_1,0) \approx \chi_{as}(1;K_1,0) \sim  \tau ^{-\gamma(1;0)}
\label{chiasising}
\end{equation}
and 
\begin{equation}
\xi(1;K_1,0) \approx \xi_{as}(1;K_1,0) \sim  \tau ^{-\nu(1;0)}
\label{xiasising}
\end{equation}
 in the critical region.  For brevity, only in this subsection we have
 set $\tau=\tau(1;0)=1-K_1/K_{1c}(1;0)$.  The crossover from the $3D$
 to the $2D$ critical behavior, as $R \rightarrow 0$, can be described
 in terms of a direct
 generalization\cite{riedel,fishmod,pft,pfeuty,fishjas,stanley} of the
 usual phenomenological scaling hypothesis valid for isotropic
 systems.  Specifically, it is assumed that, for sufficiently small
 $\tau$ and $R$, the scaling form of the singular part $
 f(1;\tau,h,R)$ of the free energy in a field $h$ embodies also the
 anisotropy parameter $R$ as follows
\begin{equation}
 f(1;\tau,h,R) \approx \tau^{2-\alpha(1;0)} F(h\tau^{-\beta(1;0)-\gamma(1;0)},
R\tau^{-\phi})
\label{freescalising}
\end{equation}
 where $F$ is a universal function.  
The exponents $\alpha(1;0)$ and $\beta(1;0)$ refer to the specific
 heat and the magnetization of the $2D$ Ising model. The quantity $\phi$,
 called {\it crossover exponent}, is universal and must
 coincide\cite{abe,suzuki,coniglio,stanley} with $\gamma(1;0)$,
  the exponent of the susceptibility of the $2D$ Ising
 model. 
 Taking two derivatives with respect to $h$ in eq.(\ref{freescalising}) 
 one obtains that  the susceptibility in zero field is given by 
\begin{equation}
\chi(1;K_1,R) \approx A^{(0)}\tau ^{-\gamma(1;0)} X^{(0)}
(B^{(0)}R\tau^{-\phi}).
\label{scalising}
\end{equation}
 where $X^{(0)}(x)$,   called universal  
{\it susceptibility   crossover-scaling function} is uniquely
defined by choosing the normalization 
$X^{(0)}(0)=\frac {dX^{(0)}(0)} {dx}= 1$.
$A^{(0)}$ and $B^{(0)}$ are non-universal scale factors. 
 Here and in what follows, a superscript zero is attached 
 to all  quantities  related
 to the 0th moment of the correlation-function.

  The scaling forms eqs.(\ref{freescalising}) and (\ref{scalising})
 provide an interpolation between the critical behaviors in $2D$
 (i.e. for $R=0$) and in $3D$, for small non-vanishing $R$, and thus
 can describe both of them. In particular, by the normalization of
 $X^{(0)}(x)$, eq.(\ref{scalising}) is consistent with the $2D$
 critical behavior eq.(\ref{chiasising}) of the susceptibility in
 $2D$.  On the other hand, the consistency of eq.(\ref{scalising})
 with the $3D$ critical behavior $\chi(1;K_1,R) \approx {\tilde
 A^{(0)}} [\tau(1;R)]^{-\gamma(1;R)}$ is achieved by assuming that
 $X^{(0)}(x)$ is singular as $K_1 \rightarrow K_{1c}(1;R)$ and, for
 $x$ in a vicinity of $\dot x$, has the structure
\begin{equation} 
X^{(0)}(x) \approx \frac{\dot X^{(0)}} {(1-x/\dot x)^{\gamma(1;R)}}
\label{xuniv}
\end{equation}
 with 
\begin{equation}
\dot x=B^{(0)}R[\tau_R(1;0)]^{-\gamma(1;0)}
\label{xdot}
\end{equation} 
 and $\tau_R(1;0)=1-K_{1c}(1;R)/K_{1c}(1;0)$.
 Due to the universality of $X^{(0)}(x)$, also the constants 
$\dot X^{(0)}$ and $\dot x$ are universal.

For small positive $R$, the solution of eq.(\ref{xdot}) yields the
reduced shift of the critical temperature of the $3D$ Ising system
with anisotropy $R$ from the critical temperature of its $2D$ limit,
which has the following asymptotic behavior
\begin{equation}
K_{1c}(1;0)/K_{1c}(1;R)-1 \sim R^{1/\phi} 
\label{tcr}
\end{equation}
with $\phi=\gamma(1;0)$.
Therefore this important result is a simple consequence of the
 crossover-scaling ansatz eq.(\ref{scalising}) and of 
 eq.(\ref{chiasising}), the critical behavior of $\chi(1;K_1,0) $.

 The validity of the extended scaling assumptions
 eqs.(\ref{freescalising}) and  (\ref{scalising}) 
can be further tested
 by a numerical HT analysis of the asymptotic behavior 
as  $\tau \rightarrow 0$
 of the successive partial derivatives 
\begin{equation}
\Xi^{(0)}_s(1;K_1,0)=
\Big( \frac{  \partial^{s}\chi(1;K_1,R)} {\partial R^s} \Big)_{R=0}
\end{equation}
 of $\chi(1;K_1,R)$ with respect to $R$,  
evaluated in the $R = 0$ limit.  The critical behavior of these quantities is
 defined by the asymptotic form
\begin{equation}
\Xi^{(0)}_s(1;K_1,0)
\approx C^{(0)}_s(1) \tau^{-\lambda_s} 
\label{gamen}
\end{equation}
 as $\tau \rightarrow 0$.  By the extended scaling hypothesis
 eq.(\ref{scalising}) the exponents of divergence $\lambda_s$ should
 satisfy the relation
\begin{equation} 
 \lambda_s =\gamma(1;0)+s\phi= (s+1)\gamma(1;0).
\label{explamb}
\end{equation}
For $s=1$,   eq. (\ref{explamb}) can also be seen
 as an immediate consequence  of the relation  
\begin{equation}
\Xi^{(0)}_1(N;K_1,0)= 
2K_1 [\chi(N;K_1,0)]^2
\label{parder}
\end{equation}
 proved in Ref.[\onlinecite{stanley}]
  for  $N$-vector models with {\it arbitrary} $N$.

For $N=1$ and $s=2,3$,  the validity of eq.(\ref{explamb}) 
 is confirmed using  the  inequalities
\begin{equation}
8 K_1^2[\chi(1;K_1,0)]^3 \ge  
\Xi^{(0)}_2(1;K_1,0)
 \ge 4 K_1^2 [\chi(1;K_1,0)]^3
\label{ineq1}
\end{equation}
\begin{equation}
48 K_1^3[\chi (1;K_1,0)]^4 \ge 
\Xi^{(0)}_3(1;K_1,0)
\ge 8 K_1^3 [\chi(1;K_1,0)]^4.
\label{ineq2}
\end{equation}
 A proof\cite{stanley} of eqs.(\ref{ineq1}) and  (\ref{ineq2})
 is based on classical correlation inequalities known to hold
 in the $N=1$ case. 
  Some generalization of  eqs.(\ref{ineq1}) and  (\ref{ineq2})
 might still be valid also for models with $N>1$.

For the lth moment of the correlation-function, one can assume 
 the validity of the extended scaling form
\begin{equation}
 m^{(l)}(1;K_1,R) \approx 
A^{(l)}\tau^{-\gamma(1;0)-l\nu(1;0)} X^{(l)}(B^{(l)}R \tau^{-\gamma(1;0)})
 \label{m2scal}
\end{equation}
As a consequence, a generalization of eq.(\ref{explamb}) is thus 
 obtained also for the exponents of divergence $\mu_s$ 
of the successive $R$-derivatives   
\begin{equation}
 \Xi^{(2)}_s(1;K_1,0)=
\Big( \frac{  \partial^{s}m^{(2)}(1;K_1,R)} {\partial R^s} \Big)_{R=0}
\end{equation}
of the second moment of the correlation-function 
$m^{(2)}(1;K_1,R)$, which are defined by the asymptotic behavior
\begin{equation}
\Xi^{(2)}_s(1;K_1,0) \approx
C^{(2)}_s(1) \tau^{-\mu_s} 
\label{m2en}
\end{equation}
 as $\tau \rightarrow 0$.
 From eq.(\ref{m2scal}) it follows that  the exponents $\mu_s $  should 
satisfy the relation
\begin{equation}
  \mu_s =2\nu(1;0)+ (s+1)\gamma(1;0). 
\label{expmu}
\end{equation}
For $s=1$, the validity of eq.(\ref{expmu}) is immediately 
proved\cite{stanley} by using 
 an analogue of eq.(\ref{parder})  for the second moment 
${m^{(2)}}(N;K_1,R)$ of the correlation-function:
\begin{equation}
\Xi^{(2)}_1(N;K_1,0)
  = 2 K_1\{ [\chi(N;K_1,0)]^2+ 2\chi(N;K_1,0) {m^{(2)}}(N;K_1,0)\}.
\label{parderm}
\end{equation}
 Notice that, as for eq.(\ref{parder}), also the validity of
 eq.(\ref{parderm}) is not limited to the $N=1$ model.

For  $s=2$ and $3$, inequalities analogous to eq.(\ref{ineq1}) and
 eq.(\ref{ineq2}) can be derived\cite{stanley} also for 
$\Xi^{(2)}_s(1;K_1,0)$, 
thus justifying eq.(\ref{m2en}).
A numerical test\cite{stanley,lambeth,liustan}
 of eqs.(\ref{gamen}),(\ref{explamb}), (\ref{m2en}) and (\ref{expmu})
 gave support to the crossover-scaling ansatz for $N=1$.

 As a final remark, let us point out that, 
 using the variables $K_2$ and $\bar R=1/R$,  
  more convenient in the $R \rightarrow \infty$ limit in which the
  system becomes an array of one-dimensional spin chains, also the following
  relations, valid for {\it arbitrary} $N$, are obtained:
\begin{equation}
\Big( \frac{\partial\chi(N;K_2,\bar R)} {\partial \bar R} \Big)_{\bar R=0}=
4K_2 [\chi(N;K_2,0)]^2 
\label{parder1d}
\end{equation}
\begin{equation}
\Big( \frac{\partial m^{(2)}(N;K_2,\bar R)} {\partial \bar R} \Big)_{\bar R=0}=
4K_2 \{ [\chi(N;K_2,0)]^2  +2\chi(N;K_2,0) m^{(2)}(N;K_2,0)\}
\label{parderm1d}
\end{equation}
 Here $\chi(N;K_2,0)$ and $m^{(2)}(N;K_2,0)$ indicate,
respectively, the susceptibility and the second moment of the
correlation-function of the anisotropic $N$-vector model for $\bar
R=0$, i.e. in $1D$.  Eqs.(\ref{parder}), (\ref{parderm}),
(\ref{parder1d}) and (\ref{parderm1d}) are quite helpful also to
validate the computation of the bivariate series expansion.

It is now also clear\cite{pfeuty,fishjas} how to compute an expansion 
of the universal susceptibility 
crossover-scaling function $X^{(0)}(x)$ in powers of $x$.
Observing that the critical amplitudes $C^{(0)}_s(1)$ in eq.(\ref{gamen}) 
 are expressed in terms of  $X^{(0)}(x)$  as
\begin{equation}
C^{(0)}_s=A^{(0)}(B^{(0)})^s \Big( \frac{d^sX^{(0)}(x)} {dx^s} \Big)_{x=0}
\end{equation}
and that the dependence on the 
non-universal quantities $A^{(0)}$ and $B^{(0)}$ disappears from the ratios
\begin{equation}
Q_s=\frac { C^{(0)}_{s-1} C^{(0)}_{s+1}} {[C^{(0)}_s]^2},
\label{ratiosr}
\end{equation}
  the  
expansion of $X^{(0)}(x)$ for small $x$ can be  written in the form
\begin{equation}
X^{(0)}(x)= 1+x+ \frac {Q_1} {2} x^2 + \frac {Q^2_1Q_2} {3!}x^3 + 
\frac {Q^3_1 Q^2_2 Q_3} {4!}x^4 +\frac {Q^4_1 Q^3_2 Q_3^2 Q_4} {5!}x^5+
\frac {Q^5_1 Q^4_2 Q_3^3 Q_4^2Q_5} {6!}x^6...
\label{xvsr}
\end{equation}
 which is universal i.e. the coefficients are independent of the
lattice structure.  The calculation of $X^{(0)}(x)$ can be numerically
extended to the whole interval $[0,\dot x]$ by Pad\`e approximants.
 This approximation can be shown to yield an asymptotic description of the
properties of $\chi(1;K_1,R)$ which, in the range of validity of the
crossover-scaling ansatz, is  consistent with those obtained 
from other approaches.
\subsection{The $N \geq 2$ models}
A slightly different formulation of the extended phenomenological
 scaling is necessary in the $N \geq 2$ cases that will be considered
 in this subsection, simply because the asymptotic relations eqs.
 (\ref{freescalising}), (\ref{scalising}), and therefore (\ref{tcr}),
 (\ref{gamen}) and (\ref{m2en}) cannot be valid anymore.  Before
 introducing this issue, it is convenient to make a brief digression to
 recall how the critical behavior in $2D$ has been
 characterized\cite{bkt} in the $N=2$ case.  As $K_1 \rightarrow
 K_{1c}(2;0)$ from below, the divergence of the correlation-length of the $2D$
 $XY$ model is dominated by an exponential singularity
\begin{equation}
\xi(2;K_1,0) \approx \xi_{as}(2;K_1,0) = 
 D\exp \big [b \tau^{-\sigma}\big ].
\label{xiasi}
\end{equation}
 In this and the following sections, for brevity, we have set 
$\tau=\tau(2;0)=1-K_1/K_{1c}(2;0)$. The universal exponent $\sigma$ is
expected to take the value $\sigma=1/2$, while $b$ is a non-universal
 positive constant.
The critical behavior of the singular part of the free energy
 is predicted to be
\begin{equation}
f_{sing}(2;K_1,0) \approx  f_{as}(2;K_1,0) =
\tilde F [\xi_{as}(2;K_1,0)]^{-2}
\label{feasi}
\end{equation}
 with $\tilde F $ a non-universal amplitude,
 while for the susceptibility one has
\begin{equation}
\chi(2;K_1,0) \approx  \chi_{as}(2;K_1,0) =
A^{(0)}\ \tau^{-\theta\sigma}\exp \big [(2-\eta(2;0))b \tau^{-\sigma}\big ]. 
\label{chiasi}
\end{equation}
 as $K_1 \rightarrow K_{1c}(2;0)$ from below. For $K_1> K_{1c}(2;0)$
 both $\xi$ and $\chi$ are infinite.  The quantity $\eta(2;0)=1/4$ is
 the exponent characterizing the large-distance behavior at
 criticality of the spin-spin correlation-function for the $2D$ $XY$
 model.  In eq.(\ref{chiasi}), the presence of a multiplicative
 correction to the leading singular behavior by a power of the
 logarithm of $\xi$ (equivalently by a power of $\tau^{-\sigma} $) and
 the value of the exponent $\theta$ are still controversial.  A
 rediscussion\cite{balog} of the renormalization group approach
 indicates that $\theta=0$, while a recent high-order HT
 study\cite{aris} and a high-precision MonteCarlo study\cite{hasen}
 support the estimate $\theta \approx 1/16$. However, the precise
 value of $\theta$ is practically irrelevant in our discussion of
 scaling.  The numerical value of the susceptibility critical
 amplitude $A^{(0)}$ depends on the value assumed for $\theta$, but it
 is also irrelevant in the determination of the crossover-scaling
 function $X^{(0)}(x)$.

 Let us now return to the scaling issue and 
make the natural assumption that the  crossover-scaling ansatz, 
 introduced in the Ising case for the singular part of the
free energy in an external field of modulus  $h$, can be simply generalized
 to the $XY$ model case as follows
\begin{equation}
  f(2;\tau,h,R) \approx [\xi_{as}(2;K_1,0)]^{-2}F(h\xi_{as}(2;K_1,0)
[\chi_{as}(2;K_1,0)]^{1/2},R\chi_{as}(2;K_1,0))
\label{freescalkt}
\end{equation}
for sufficiently small positive $R$ and $\tau$.  Taking two derivatives
 with respect to the field, we get the generalized scaling form for the
 susceptibility in zero field
\begin{equation}
\chi(2;K_1,R) \approx \chi_{as}(2;K_1,0) X^{(0)}
\Big(B^{(0)}R\chi_{as}(2;K_1,0)\Big)
\label{scalkt}
\end{equation}
 Here $X^{(0)}(x)$ is a universal
 crossover-scaling function, that can
 be uniquely defined assuming that $X^{(0)}(0)=\frac {dX^{(0)}(0)}
 {dx}= 1$. $B^{(0)}$ is a non-universal scale
 factor.  

  The generalized crossover-scaling forms eqs. (\ref{freescalkt})
 and (\ref{scalkt}) are immediately shown to reduce
  to eqs. (\ref{freescalising}) and (\ref{scalising}) 
 in the $N=1$ case, by using eqs.(\ref{chiasising}) and (\ref{xiasising})
 and the scaling laws. However,  
 for $N = 2$, these forms have the additional virtue of  correctly 
allowing for the fact that the
 critical singularities in $2D$ are not power-like, since the exponents 
$\gamma(2;R)$ and $\nu(2;R)$ (as well as $\alpha(2;R)$ and $\beta(2;R)$) 
 are ill-defined in the $R \rightarrow 0$
 limit and  also no crossover exponent $\phi$  exists.
Of course, the scaling forms eq.(\ref{freescalkt}) and (\ref{scalkt})
 can as well be written exclusively in terms of $\xi$ thanks 
to eq. (\ref{chiasi}).

  By the normalization of $X^{(0)}(x)$, the scaling form
 eq.(\ref{scalkt}) is consistent with the $2D$ critical behavior,
 since for $R=0$  and $\tau \rightarrow 0$, 
one has $\chi(2;K_1,0) \approx \chi_{as}(2;K_1,0) $.
   On the other hand, for small but non-vanishing
 $R$, the conventional $3D$ critical behavior $\chi(2;K_1,R) \sim \tilde A
 \tau^{-\gamma(2;R)}$ as $\tau(2;R) \rightarrow 0$, is recovered by
 assuming that $X^{(0)}(x)$ has the same singularity structure as in
 eq.(\ref{xuniv}), i.e.  that $X^{(0)}(x) \approx \frac{\dot X^{(0)}}
 {(1-x/\dot x)^{\gamma(2;R)}}$, when $x$ is in a neighborhood of  
 $\dot x$, with
\begin{equation}
\dot x=B^{(0)}R\chi_{as}(2;K_{1c}(2;R),0). 
\label{aslog}
\end{equation}
 As  in the $N=1$ case, 
by solving eq.(\ref{aslog}), we can obtain
 the small-$R$  asymptotic behavior of the reduced 
critical-temperature shift of the anisotropic $3D$ model 
from its $2D$ limiting value 
\begin{equation} 
K_{1c}(2;0)/K_{1c}(2;R)-1 \approx  V/[{\rm ln} (R/W)]^2.
\label{tcr2}
\end{equation}

Here $V=[2-\eta(2;0)]^2 b^2$ and $W= \dot x/BA^{(0)}$ are
non-universal constants. The higher-order corrections to the leading
behavior in eq.(\ref{tcr2}), are also expressed in terms of inverse
powers of $|{\rm ln} R|$ and possibly of ${\rm ln|ln} R|$, depending
on the value of the exponent $\theta$.  It is interesting to point out
that the original argument\cite{ktcro,hikami} for eq.(\ref{tcr2})
relied on approximate renormalization group ideas, rather than being a
simple consequence of the generalization of the crossover-scaling
ansatz and of the exponentially-singular critical behavior of the $XY$
model.

As a further immediate  consequence of our scaling assumption 
eq.(\ref{scalkt}),  
 the  divergence  of the successive $R$-derivatives of the 
 susceptibility  turns out to have the structure 
\begin{equation}
\Xi^{(0)}_s(2;K_1,0)
\approx  C^{(0)}_s(2) [\chi_{as}(2;K_1,0)]^{s+1} 
\sim \tau^{\theta \sigma(s+1)} exp[(2-\eta(2;0))b(s+1)\tau^{-\sigma}]
\label{gamenweak}
\end{equation}
 as $\tau \rightarrow 0$, which can be seen as  a natural generalization 
of eq.(\ref{gamen}). For $s=1$, the validity of
 eq.(\ref{gamenweak}) is an obvious consequence of eq.(\ref{parder}).
  For $s=2$ and $3$, it might follow from some extension of the
 inequalities eq.(\ref{ineq1}) and eq.(\ref{ineq2}). Eq.(\ref{gamenweak})
 will be numerically tested in subsection IV C, by studying the behavior of  
$\Xi^{(0)}_s(2;K_1,0)$ as $K_1 \rightarrow K_{1c}(2;0)$, for the 
first six values of $s$.

 By assuming for the second moment $ m^{(2)}(2;K_1,R) $ of the
 correlation-function the crossover-scaling form
\begin{equation}
m^{(2)}(2;K_1,R) \approx m_{as}^{(2)}(2;K_1,0)
X^{(2)}\Big( B^{(2)}R \chi_{as}(2;K_1,0)\Big)
\label{mlweak}
\end{equation}
 where 
$ m_{as}^{(2)}(2;K_1,0)=4[\xi_{as}(2;K_1,0)]^2\chi_{as}(2;K_1,0)$, 
the  analogue of eq.(\ref{m2en}) is obtained  
\begin{equation}
\Xi^{(2)}_s(2;K_1,0)
 \approx C^{(2)}_s(2)  [\xi_{as}(2;K_1,0)]^2 [\chi_{as}(2;K_1,0)]^{s+1}
\label{gamemweak}
\end{equation}
By eq.(\ref{parderm}), this equation is certainly valid for $s=1$ and 
we shall suppose that it is true also for $s>1$. Also eq.(\ref{gamemweak})
 will be tested numerically in subsection IV C, by the same method used 
for eq.(\ref{gamenweak}).

The HT expansions of $\Xi^{(0)}_s(2;K_1,0)$ and $\Xi^{(2)}_s(2;K_1,0)$
 can be immediately read from our tables of the series coefficients 
 of  $\chi(2;K_1,K_2)$ and  $m^{(2)}(2;K_1,K_2)$, respectively.

 The expansion in powers of $x$
 of the universal scaling function $X^{(0)}(x)$ can be computed 
by the  same procedure already outlined for the $N=1$ case. 

A generalized scaling assumption can be made also to describe the crossover
 of the $3D$ system to a set of one-dimensional non-interacting $XY$ chains as 
$ R \rightarrow \infty$. It is now convenient to shift to the variables
 $K_2$ and $\bar R=1/R$.
One has simply to observe that in this case  
$K_{2c}(2;\bar R) \rightarrow \infty$
as $\bar R \rightarrow 0$  and that, 
at criticality, the divergence of the susceptibility in one dimension is
\begin{equation}
\chi(2;K_2,0)\approx \chi_{as}(2;K_2,0)= \bar A^{(0)} K_2^2
\label{chias1d}
\end{equation}
Then it is natural to  introduce a universal susceptibility scaling function 
$\bar X^{(0)}(x)$, normalized like $ X^{(0)}(x)$ and with the same singularity
 structure and    assume that
\begin{equation}
\chi(2;K_2,\bar R) \approx \chi_{as}(2;K_2,0) \bar X^{(0)}
\Big(\bar B^{(0)} \bar R\chi_{as}(2;K_2,0)\Big).
\label{scal1d}
\end{equation}
 It follows  that 
 \begin{equation} 
K_{2c}(2;\bar R) \sim \bar R^{-1/2}
\label{tcr2b}
\end{equation}
 as $\bar R \rightarrow 0$.  Moreover, predictions for the $\bar
 R$-derivatives of the susceptibility, similar to those mentioned
 above for the $R$-derivatives, are easily  obtained along the same lines.

 Let us finally point out that generalized crossover-scaling assumptions
of the same form as eqs.(\ref{freescalkt}) and (\ref{scalkt}) can be
written down also for the $N$-vector spin models with $N>2$. The main 
difference is that for these models, $K_{1c}(N;R) \rightarrow \infty $ as 
$ R \rightarrow 0 $ and that, in $2D$, the critical divergence of the
correlation-function moments and of the correlation-length is
exponential in $K_1$.  As a result, we can conclude that
 for small $R$, 
 we have $K_{1c}(N;R) \sim |{\rm ln}R| $, by the same scaling arguments 
used above.
 Here, we shall not further investigate the models with
$N>2$, but only note that a general discussion\cite{bcoen2d} of the
numerical difficulties of the HT-expansion approach to the $N$-vector
models in $2D$ suggests that a HT study of the crossover might also
meet with similar problems caused by the unphysical singularities in
the complex inverse-temperature plane revealed by a large-$N$
study\cite{bcoinf}.
\section{A numerical analysis}
We shall now turn to an analysis of our HT expansions to study the
 crossover behavior (\ref{tcr2}) of the critical inverse-temperature
 $K_{1c}(2;R)$ as $R \rightarrow 0$ or $R \rightarrow \infty$, 
to check the universality with
 respect to $R$ of the critical exponents of the anisotropic system,
 to test the validity of the consequences eqs. (\ref{tcr2}), 
(\ref{gamenweak}), (\ref{gamemweak})
 and (\ref{tcr2b}) of the crossover-scaling assumptions
 eqs.(\ref{scalkt}), (\ref{mlweak}), (\ref{scal1d}) 
  and finally to obtain an approximation of the
 susceptibility scaling function $X^{(0)}(x)$.
\subsection{ Estimates of the critical temperature as function of $R$ }
Let us first study the behavior of the susceptibility $\chi(2;K_1,R)$
as function of $K_1$ at fixed values of $R$, to determine the critical
locus. We shall later use these results to bias the computation of the
critical exponents of the susceptibility and of the correlation-length
and verify that they satisfy the universality hypothesis, as long as
$R >0 $.  For simplicity, we have analyzed the behavior of our
expansions as functions of $K_1$ at fixed values of $R$ (or as
functions of $K_2$ at fixed values of $\bar R$), using the
conventional {\it single-variable} methods\cite{Gutt} of series
analysis, namely PAs or inhomogeneous differential approximants (DAs).
It may be helpful to recall that in the DA approach a
(single-variable) power series is resummed by expressing it as the
solution of a linear (first- or higher-order) differential equation
with polynomial coefficients and inhomogeneous term, appropriately
defined in terms of the coefficients of the given series. 
We believe that taking advantage of the large number of series
  coefficients presently available, also more complex methods
  of series analysis, like multivariate PAs\cite{alabiso} or 
  partial-differential approximants\cite{fisherkerr}, become now worth
  exploring. However, we have not yet thoroughly pursued these approaches.

In our computation of the critical temperature, for $R > 0.05$, we
  have used  second-order DAs and have considered the class of
   $[k,l,m;n]$ DAs restricted by the conditions: $13 \leq
  k+l+m+n \leq 19$ with $k \geq 3; l \geq 3; m \geq3$.  Among these,
  we have selected the DAs with the additional properties of being defect-free,
  i.e. having sufficiently isolated physical singularities, and of
  being near-diagonal, i.e. such that $|k-l|, |l-m| \leq 2$, $1< n <
  4$. Finally, we have not ``biased'' the DAs, i.e. we have not discarded
 the approximants yielding critical exponents  with values outside
 appropriate limits. 
 Here we are using the standard notation in which $k,l,m,n$
  denote the degrees of the polynomial coefficients and of the
  inhomogeneus term of the differential equation defining the DA.
  These rather technical specifications are given only to make our
  results completely reproducible. However, we have always made sure
  that our final estimates, within a fraction of their uncertainties,
  are essentially independent of the precise definition of the DA
  class examined.  At a given value of $R$, our central estimate of
  $K_{1c}(2;R)$ is the sample mean of the locations of the
  singularities in this class of DAs, taken after dropping evident
  outliers.  A small multiple of the spread of the reduced sample is
  taken as an estimate of the uncertainty. Possible residual
  unsaturated trends of the central estimates, as the number of series
  coefficients used in the calculation is increased, have been
  accounted for by attaching generous error bars to the final
  estimates. Sometimes we have made small corrections of these
  central estimates (always well within the uncertainties) suggested
  by a comparison with the sequence of Zinn-Justin
  modified-ratio\cite{Gutt,bcisiesse} estimates.

  Coming to the numerical results, let us first note that, for $R=1$, 
  our estimate $K_{1c}(2;1)=0.22710(3)$ compares well both with our
  older estimate\cite{bcoen} $K_{1c}(2;1)=0.227095(10)$ obtained from
  HT expansions of order 21 for the isotropic $3D$ $XY$ system and
  with the much more precise\cite{deng} recent determination
  $K_{1c}(2;1)=0.2270827(5)$, obtained from a high-accuracy MonteCarlo
  simulation.

\begin{figure}[tbp]
\begin{center}
\leavevmode
\includegraphics[width=3.37 in]{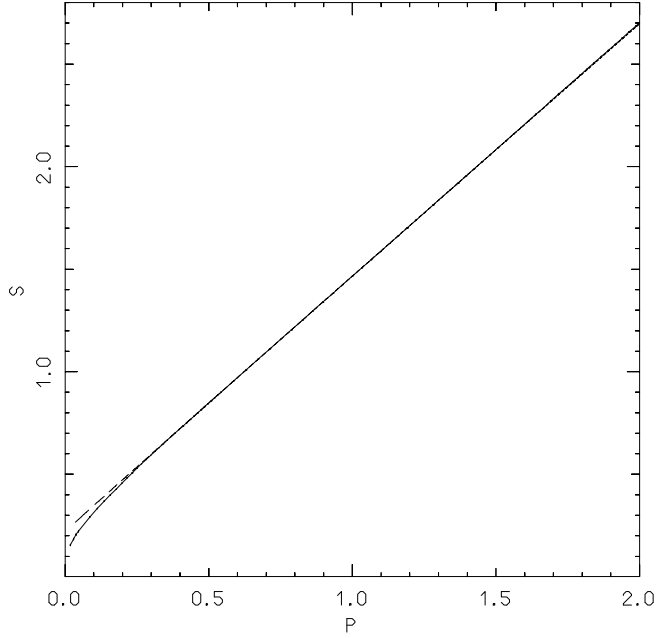}
\caption{\label{anisf2y2} The quantity  $S=K_{1c}(2;0)/K_{1c}(2;R)-1$,  which
 represents the reduced shift of the critical temperature of the
 system with anisotropy $R$ from its $2D$ limit, 
  is plotted vs $P=R^{2/3}$.  A continuous line
 interpolates our estimates, whose error bars are smaller than the
 width of the line, except for very small $R$.  The 
 dashed line (hardly visible except for small $R$ ),  which is 
superimposed to the continuous one,
  is  the result of a fit of the expression $f(R)=aR^g+c$ to our data for
 $0.05< R < 3.4 $. The values of the fit parameters are $a \approx
 1.245$, $g \approx 0.661$ and $c \approx 0.221$.  }
\end{center}
\end{figure}

 The determination of the line of critical points $K_{1c}(2;R)$ for
  very small values of $R$, as required for a numerical test of the
  predicted crossover behavior eq.(\ref{tcr2}), is a delicate task
  both by series study and by simulation.  The progressive decoupling
  of the horizontal layers as $R \rightarrow 0$, makes a reliable
  extrapolation of the susceptibility to its genuinely $3D$ critical
  behavior possible only from a closer and closer vicinity of the
  critical point. (See also the comments to Fig.\ref{anis_f22} in the
  next subsection.)  As a consequence, the precision of the estimates
  of the location of the critical point and of the critical exponents
  tends to deteriorate as $R \rightarrow 0$.  We have observed that,
  for $R \lesssim 0.05$, a reasonable increasing behavior of
  $K_{1c}(2;R)$ is still obtained simply by assuming what will be
  actually borne out by our analysis, namely that the critical
  exponent of the susceptibility is universal, to a good
  approximation, all along the critical locus for $R>0$ and thus by
  imposing also a ``weak'' bias on this exponent when computing
  $K_{1c}(2;R)$. This simply amounts to discard from the sample of our
  data the critical temperature estimates obtained from DAs whose
  exponent at the singularity differs by more than $10\%$ from the
  expected value of $\gamma(2;1)$. Even after this simple improvement
  of the analysis, for $R < R_{min} \approx 0.0015$ , our expansions
  do not anymore seem to be long enough to locate the critical point 
 with acceptable  accuracy. Therefore we shall not report estimates
  for $R< R_{min}$. Our results for the reduced shift
  $S=K_{1c}(2;0)/K_{1c}(2;R)-1$ of the critical temperature from its $2D$
 limiting value
  are plotted vs $R^{2/3}$ in Fig.\ref{anisf2y2}.  The value
  $K_{1c}(2;0)=0.56000(5)$ of the critical inverse-temperature of the
  $2D$ $XY$ model on the square lattice has been taken from
  recent\cite{bperxy,aris,hasen} high-precision studies.

 A short list of our  
 numerical DA estimates  of $K_{1c}(2;R)$ for
  $0.005 \leq R \leq 3.4$ 
 can be found in Table \ref{tabella}. 

 It should  be noticed  that the critical curve must be separately
  symmetric under the transformations $ K_1 \rightarrow -K_1$ and $
  K_2 \rightarrow -K_2$.  Therefore the ferromagnetic phase diagram,
  so far represented, should be completed by the remaining branches of
  the critical locus.

In Fig.\ref{anisf2y2} we have also shown that $S$ has a very simple
dependence on $R$, valid to a high accuracy over a wide range of
intermediate-large (but not too large) values of $R$. The expression
$f(R)=aR^g+c$ with $a \approx 1.245$, $g \approx 0.661$ and $c \approx
0.221$ interpolates quite accurately our data points for $S$ in the interval
$ 0.025 <R < 2.9$, visibly departing from them only for small $R$,
i.e. in the region where the crossover is expected to occur. The value
of the exponent $g$ justifies our choice to plot $S$ vs $R^{2/3}$.  
It would be interesting to look for an analytic explanation
of this purely empirical remark.

 It is interesting to remember that a not very different $R^{4/7}$ law was
conjectured\cite{weber} to describe the $R$-dependence of the critical
temperature over the region $0.01 < R< 0.7$, while for $R>0.7$, a
linear law was proposed.  This suggestion was based on a
self-consistent mean-field approximation, which is probably not very accurate
for small $R$ and is certainly inaccurate for intermediate-large $R$.

\begin{figure}[tbp]
\begin{center}
\leavevmode
\includegraphics[width=3.37 in]{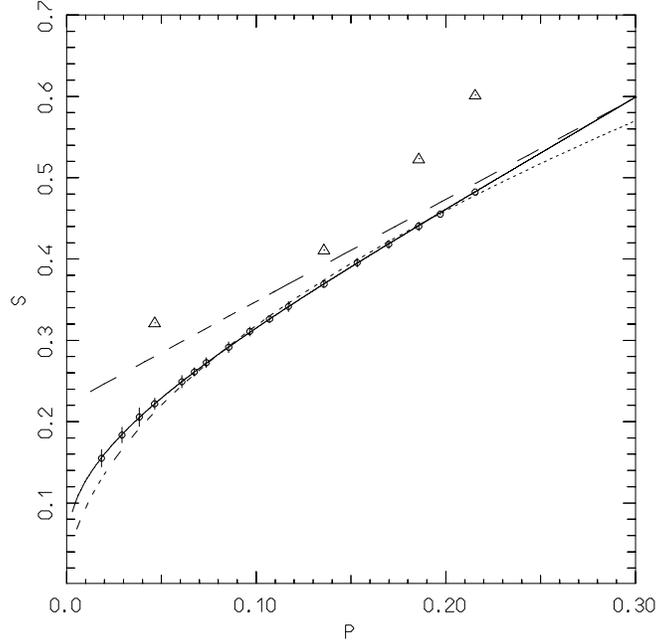}
\caption{\label{anisf3sq} A blow-up of the lower left corner of 
 Fig.\ref{anisf2y2}, showing the plot of the reduced critical-temperature shift
 $S=K_{1c}(2;0)/K_{1c}(2;R)-1$  vs  $P=R^{2/3}$. The
 continuous line is a fit of the expression $ V/[{\rm ln} (R/W)]^2$ to
 our estimates of $S$ (circles) for
 $0.005<R<0.15$. The values of the parameters are
 $V \approx 11.34$ and $W \approx 12.7$.
 The short-dashed line is the result of a fit of the expression
 $\tilde f(R)=aR^{g'}$, with $a \approx 1.08$ and 
$g' \approx 0.354$, to the same set of data.  The long-dashed line is the
 result of the same fit as in Fig.\ref{anisf2y2} 
of the expression $f(R)=aR^g+c  $, with parameters
 $a \approx 1.245$, $g  \approx0.661$ and $c \approx 0.221$, 
to all data for $0.05 \leq R \leq 3.4 $.  
The triangles are  small-$R$
 estimates of $S$ taken from the simulation of Ref.[\onlinecite{chui}] }
\end{center}
\end{figure}

Let us now turn to the small-R region to study the features of the
 crossover behavior.  Fig.\ref{anisf3sq} is a blow-up of the lower
 left corner of Fig.\ref{anisf2y2} showing our estimates of the
 critical temperature for $R<0.13$.  We can notice that they clearly
 depart from the simple $R^{2/3}$ behavior indicated by a long-dashed
 line.  A continuous line indicates the result of a fit of the
 asymptotic expression $ V/[{\rm ln} (R/W)]^2$ of eq.(\ref{tcr2})
 predicted by the crossover-scaling theory to describe the behavior of
 the reduced temperature shift in the crossover region.  Restricting
 the fit to the critical-temperature estimates which fall within the
 window $(0.0025,R_{max}=0.1)$, we determine the values of the
 parameters $V \approx 11.34$ and $W \approx 12.7$.  In the same
 figure, for comparison, we have reported also some old MonteCarlo
 estimates\cite{chui} of $K_{1c}(2;R)$ in the small $R$ range.  In
 spite of their order-of-magnitude agreement with our series
 estimates, these simulation data seem to suggest a qualitatively
 different behavior as $R \rightarrow 0$. We must therefore suppose
 that they are affected by large errors (unfortunately not assessed in
 Ref.[\onlinecite{chui}]) from underestimated finite-size effects,
 because the volume of the simulated system was small (at most $20^3$)
 and no finite-size-scaling analysis was performed.  A more recent
 simulation\cite{kawa}, also carried out with $20^3$ sized systems,
 and extending to much smaller values of $R$, is likely to suffer from
 similar problems, although its results seem to show a better
 agreement with ours.

 Some remarks have to be made on Fig.\ref{anisf3sq}. Firstly, one
  should be aware that a logarithmic behavior is quite difficult to
  identify numerically by using data which refer to only a two-decade
  variation of the independent variable, since, over a restricted
  range, it can also be well represented as a power-law behavior with
  a small exponent.  Moreover the choice of the window of values of $R$
 to be studied, is
  a delicate issue, because the upper end of the window should be
  sufficiently small that both the crossover-scaling assumptions and
  the asymptotic behavior eq.(\ref{tcr2}) apply with small
  corrections, while the lower end should be sufficiently large that
  our estimates of the critical temperature are not too uncertain.  In
  our case the value of $R_{max}$ can be varied by a factor of two or
  more, still obtaining good fits of the same functional form, with
  not very different values of the parameters $V$ and $W$.  Finally,
  we observe that the best-fit value of $V$ is not very different from
  its expected value $(2-\eta(2;0))^2 b^2 \approx 9.5 $, while the
  value of $W$ is much larger. It is reasonable to interpret these results
 as an indication that $R_{min}$ is
  still too large and therefore the higher-order corrections to the
  asymptotic form eq.(\ref{tcr2}), suppressed only
  by inverse powers of ${\rm |ln}R|$, are still important,  so that $V$ and
  $W$ can only be effective parameters.  To conclude, in spite of the
  notable extension of the HT expansions we have analyzed, these results still
  can give only a suggestive indication that our estimates of the critical
  temperature for small $R$ are compatible with the predicted
  asymptotic behavior eq.(\ref{tcr2}).

On the other hand, one might wish to describe also this crossover behavior 
by a power law and fit an expression $\tilde f(R)=aR^{g'}$,  
of the same form as eq.(\ref{tcr}) valid in the $N=1$ case,  to the same
small-$R$ sample of our data points.  We then find  the following
values of the parameters: $a \approx 1.08$ and $g' \approx 0.354$.
 We have displayed also this fit in Fig.\ref{anisf3sq}.
Thus, if  for consistency we  describe also 
the critical behavior of the $2D$ $XY$ model 
 in terms of conventional power laws, this result  suggests an unusually
 large susceptibility
 exponent,   i.e. $\gamma(2;0)=1/g' \approx 2.8$.
 We can also observe that, subdividing the small-$R$ range
 into two intervals, this kind of fit would yield a smaller exponent $g'$
 in the leftmost interval.  This suggests that even smaller values of
 the exponent $g'$ (and thus larger values of the susceptibility exponent) 
are likely to be found, if it were possible to
 further reduce both ends of the window of values of $R$ under
 consideration, for example by using future significantly longer HT
 expansions.
It is then reasonable to believe that
 this power-law crossover is only apparent, because the exponent is
$R$-dependent 
and vanishes as $R \rightarrow 0$, and hence  that we  are actually 
representing a genuine logarithmic behavior as a power-law behavior.

\begin{figure}[tbp]
\begin{center}
\leavevmode
\includegraphics[width=3.37 in]{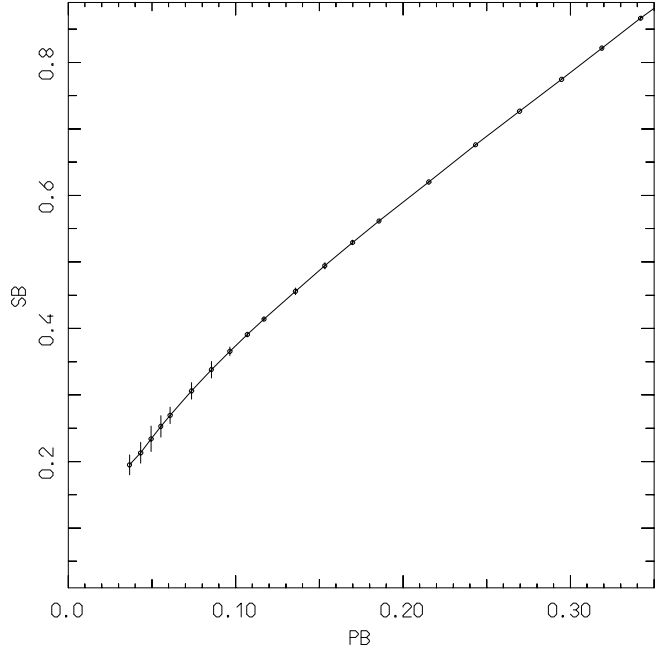}

\caption{\label{anisf3_1suR} 
A plot of $SB=1/2K_{2c}(2;\bar R)$ (circles)
 vs $PB=\bar R^{2/3}$, in the small $\bar R$ region. A continuous line
 interpolates our estimates.  }
\end{center}
\end{figure}

 In Fig.\ref{anisf3_1suR}, we have plotted
 $(2K_{2c}(2;\bar R))^{-1}$ vs $\bar R^{2/3}$. The 
behavior of the curve as $\bar R \rightarrow 0$ also gives some support
 to the validity of eq.(\ref{tcr2b}) and therefore it confirms
 the generalized scaling approach to the crossover from $3D$ to $1D$.
\subsection{ Universality of the critical exponents with respect to $R$}
 Unfortunately, so far we have computed expansions only for the $sc$
 lattice structure and therefore our verification of the universality
 of certain quantities does not include their lattice independence,
 but is limited to a test of their independence on $R$.

\begin{figure}[tbp]
\begin{center}
\leavevmode
\includegraphics[width=3.37 in]{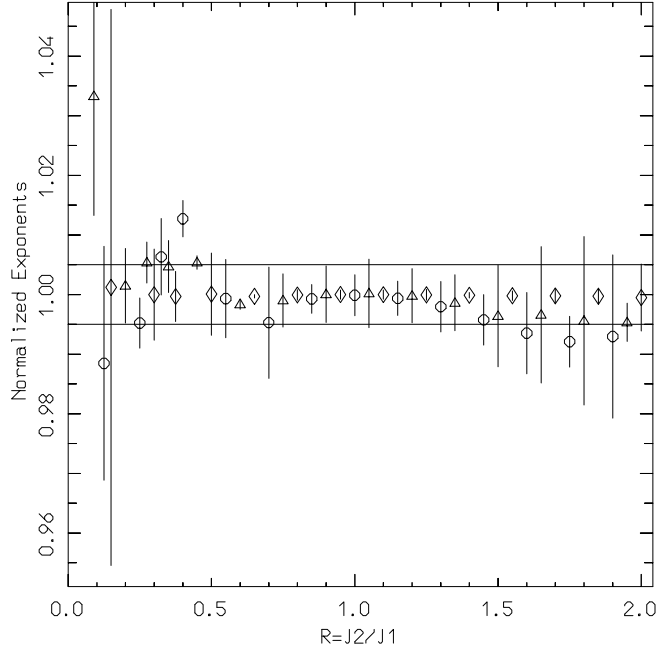}
\caption{\label{anisf21} Estimates of the exponents $\gamma(2;R)$
 (circles), $\nu(2;R)$ (triangles) and of the ratios
 $\nu(2;R)/\gamma(2;R)$ (rhombs),  plotted vs the anisotropy parameter $R$.
 All these quantities are computed by DAs biased with the critical temperature 
 and they are normalized to our estimates of
 their values at $R=1$.
   The horizontal lines are bands of $0.5\%$ deviation from our
 estimates of $\gamma(2;1)$, $\nu(2;1)$ and $\gamma(2;1)/\nu(2;1)$. }
\end{center}
\end{figure}

  In the range $R > 0.05$, in which our computation of the critical
 temperature was completely unbiased, we have also evaluated both the
 critical exponents $\gamma(2;R)$ and $\nu(2;R)$ by using second-order
 DAs, chosen in the above specified class and biased with our estimates
 of $K_{1c}(2;R)$.  We have also determined, by simple first-order
 DAs, the ratio of the logarithmic derivatives of $m^{(2)}(N;K_1,
 R)/K_1$ and $\chi(N;K_1, R)$, whose value at $K_{1c}(2;R)$ yields the
 ratio $ \nu(2;R)/\gamma(2;R) $. In Fig.\ref{anisf21}, we have plotted
 vs $R$ our estimates of these exponents and of their ratio normalized
 to our estimates of $\gamma(2;1)=1.328(10) $, $\nu(2;1)=0.679(8)$ and
 $\nu(2;1)/\gamma(2;1)= 0.5099(6)$, respectively.  The central values
 of our estimates of the normalized exponents and of the exponent
 ratio are very near to unity and fairly independent of $R$ within
 $\approx 0.5\%$, over a range wider than that shown in the figure,
 except for very small or large $R$, due to the expected crossover.
 These results support the universality with respect to $R$ of the critical
 behavior for the anisotropic $3D$ model.  It is fair to note that,
 while the deviations of our exponent estimates for the anisotropic
 system from those for the $R=1$ system are uniformly very small, our
 estimates of $\gamma(2;1)$ and $\nu(2;1)$ are larger, by $\lesssim
 1\%$, than the most recent\cite{bcisiesse,xypisa} estimates
 $\gamma(2;1)=1.3178(2)$ and $\nu(2;1)=0.67155(27)$.  On the contrary,
 the deviation of our estimates of the exponent ratio from its recent
 determination $\nu(2;1)/\gamma(2;1)= 0.5096(3)$ is quite
 small. Substantially longer series or improved methods of exponent
 determination\cite{bcisiesse,bplif,xypisa} might be necessary to
 remove the discrepancy in the exponent estimates, which certainly can
 be ascribed to the residual influence of the corrections to scaling.

 In Fig.\ref{anis_f22} we have plotted, for various fixed values of
 $R$, the {\it effective exponent}\cite{kouv,riedel} $\gamma_{eff}(2;K_1,R)$ 
of the susceptibility
\begin{equation}
 \gamma_{eff}(2;K_1,R)=-\frac{d {\rm ln} \chi(2;K_1,R)} {d {\rm ln} \tau(2;R)}.
\label{gameffect}
\end{equation}
 Roughly speaking, this quantity represents the {\it local} value of
 the critical exponent which would be inferred by a measure of the
 susceptibility in a neighborhood of $K_1$ and only in the critical
 limit it coincides with the asymptotic critical exponent.  By showing
 how the local critical behavior changes, this quantity is helpful to
 describe the crossover phenomena.  The curves shown in the figure are
 obtained simply by forming the highest-order, defect-free, diagonal
 or near-diagonal PAs of the HT expansion of the right-hand side of
 eq.(\ref{gameffect}).  The figure illustrates the narrowing down of
 the genuinely $3D$ critical region as $R \rightarrow 0$, that was
 already discussed in the preceding subsection. More precisely, when
 $R$ is not too small, the effective exponent is approximately
 independent of the temperature and stays close to its $3D$ asymptotic
 value. On the other hand, going to sufficiently small $R$, the
 effective exponent becomes increasingly temperature-sensitive: in
 most of the HT temperature range it takes rather large values, which
 somehow reflect the exponential $2D$ critical singularity, and only
 as $K_1$ gets very close to the critical value, it approaches the
 $3D$ asymptotic value.

\begin{figure}[tbp]
\begin{center}
\leavevmode
\includegraphics[width=3.37 in]{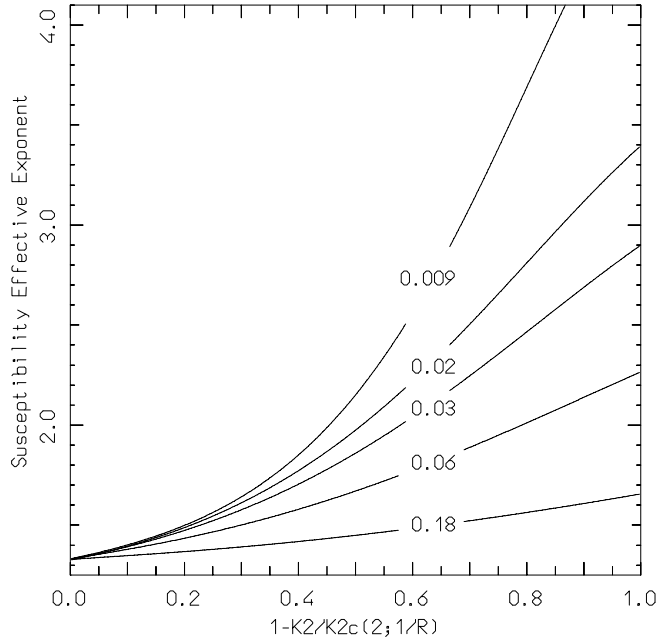}
\caption{\label{anis_fRb} 
The effective exponent
$\gamma_{eff}(2;K_2,\bar R)$, computed by PAs, for the susceptibility of
the anisotropic system vs $\tau(2;\bar R)=1-K_2/K_{2c}(2;\bar R)$ 
for various fixed
values of  $\bar R$ indicated on the corresponding curves.  }
\end{center}
\end{figure}

 In Fig.\ref{anis_fRb} we have plotted, for various fixed values of
 $\bar R$, the effective exponent $\gamma_{eff}(2;K_2,\bar R)$ of the
 susceptibility, defined in strict analogy with eq.(\ref{gameffect})
 to describe the crossover from $3D$ to $1D$.  Also in this case, the
 curves indicate a transition from a region of very high values of the
 effective exponent, reflecting the critical divergence of the
 susceptibility of the $1D$ system, to the asymptotic region where the
 $3D$ critical behavior is attained.
\subsection{Critical behavior of $\Xi^{(0)}_0(2;K_1,0))$. An approximation 
of the  susceptibility crossover-scaling function $X^{(0)}(x)$.}
 As  sensitive and specific indicators to test the 
critical behavior eq.(\ref{gamenweak}) 
of the successive $R$-derivatives of the  susceptibility 
$\Xi^{(0)}_s(2;K_1,0)$, we have formed the normalized ratios of their
logarithmic derivatives 
\begin{equation}
G^{(0)}_s(K_1)= \frac {1} {s+1} \frac {d {\rm ln}[\Xi^{(0)}_s(2;K_1,0)]/dK_1}
{d {\rm ln}[\Xi^{(0)}_0(2;K_1,0)]/dK_1}
\end{equation}
 If eq.(\ref{gamenweak}), showing the critical behavior of
$\Xi^{(0)}_s(2;K_1,0)$, is valid, the quantities $G^{(0)}_s(K_1)$ are
expected to tend to unity, as $\tau \rightarrow 0$, independently of
$s$.  This result is obvious for $s=1$, thanks to the exact relation
eq.(\ref{parder}), but for $s>1$ it ought to be numerically tested.
We have used near-diagonal first-order DAs of the HT expansions of
$G^{(0)}_s(K_1)$ to estimate the limits of these quantities as $K_1
\rightarrow K_{1c}(2;0)$.  Our results, summarized in the Table
\ref{tabella2}, support the validity of the asymptotic relation
eq.(\ref{gamenweak}) to a good approximation when $s=2,3,4,5,6$ and
therefore lend support to the validity of the generalized
crossover-scaling assumption eq.(\ref{scalkt}).  In general, it should
not be surprising that the precision of these, as well as of the
following estimates, is smaller than that reported in previous studies
of much shorter series in the $N=1$ case, in which the critical
singularities are power-like, because of the more complex structure of
the critical singularity in the $2D$ $XY$ model.  One should also
notice that the number of nontrivial expansion coefficients of the
$R$-derivatives of the moments of the correlation-function and hence
the precision of the numerical approximation for the quantities
related to them, decreases as $s$ increases. Therefore these tests are
less reliable for values of $s$ larger than those examined here.

A completely parallel study of the indicator $G^{(2)}_s(K_1)$, a
strict analogue of $G^{(0)}_s(K_1)$ for the quantity 
$\frac{\Xi^{(2)}_s(2;K_1,0) }{[\xi(2;K_1,0)]^2}$, which can be associated 
to the $R$-derivatives of the second moment of the correlation-function,
also leads to results in agreement with eq.(\ref{gamemweak}), albeit
slightly less precise, because of the known slower convergence of the
second moment expansion.  Thus also the validity of eq.(\ref{mlweak})
is confirmed. Our estimates of the limits of the quantities
$G^{(2)}_s(K_1)$ as $K_1 \rightarrow K_{1c}(2;0)$ are also reported in
the Table \ref{tabella2}.

Let us now turn to the study of the universal crossover-scaling
function $X^{(0)}(x) $ of the susceptibility, following the
procedure\cite{pfeuty,fishjas} described above in detail for the $N=1$
case at the end of subsection III A.  As a first step, we have to
determine the critical amplitudes $C^{(0)}_s(2)$ of
$\Xi^{(0)}_s(2;K_1,0)$, defined in eq.(\ref{gamenweak}) by the limit
of the {\it effective amplitudes} 
$C^{(0)}_s(2;K_1,0)=\Xi^{(0)}_s(2;K_1,0)/\chi^s(2;K_1,0) $ as $K_1
\rightarrow K_{1c}(2;0)$.  The amplitudes $C^{(0)}_s(2)$ are then used
to form the universal ratios $Q_s$, defined by eq.(\ref{ratiosr}).
Alternatively, the ratios $Q_s$ can also be determined, directly and
perhaps more accurately, by extrapolating the HT expansions of the {
\it effective ratios}
\begin{equation}
Q_s(2;K_1,0)= \frac{\Xi^{(0)}_{s-1}(2;K_1,0)\Xi^{(0)}_{s+1}(2;K_1,0)}
 {[\Xi^{(0)}_{s}(2;K_1,0)]^2}
\label{effrat}
\end{equation} 
to $K_1=K_{1c}(2;0)$. The estimates so obtained for the universal quantities 
 $Q_s=\lim_{K_1 \to K_{1c}} Q_s(2;K_1,0) $ are reported in Table
\ref{tabella4}. 

 The coefficients of the small-$x$ expansion of $X^{(0)}(x)$
are finally expressed in terms of the  $Q_s$, as
indicated in eq.(\ref{xvsr})
\begin{equation}
X^{(0)}(x)=1+x+0.792(3)x^2+0.600(5)x^3+0.44(1)x^4+0.32(1)x^5+0.23(2)x^6+
0.16(3)x^7+...
\label{xdx}
\end{equation}
Having assumed that $X^{(0)}(x)$ is singular at $\dot x$ and has the form
$X^{(0)}(x) \approx \frac{\dot X^{(0)}} {(1-x/\dot x)^{\gamma(2;R)}}$ in 
 a vicinity of $\dot x$, we can give a reasonably good estimate of $\dot x$ by
locating the nearest pole of the highest-order, defect-free,
near-diagonal PAs of the expansion of $[X^{(0)}(x)]^{1/\gamma(2;R)}$.
Choosing $\gamma(2;R)=1.3178(2)$ and allowing for the uncertainties of
the expansion coefficients in eq.(\ref{xdx}) and the spread of the PA
singularities, we can estimate $\dot x=1.475(15)$ and $\dot
X^{(0)}=1.154(15)$. A direct estimate of $\dot x$, by extrapolating to
$R=0$ the argument $BR\chi_{as}(2;K_{1c}(2;R),0)$ of the scaling
function, fails to yield a more accurate result because of the
extrapolation uncertainties.

\begin{figure}[tbp]
\begin{center}
\leavevmode
\includegraphics[width=3.37 in]{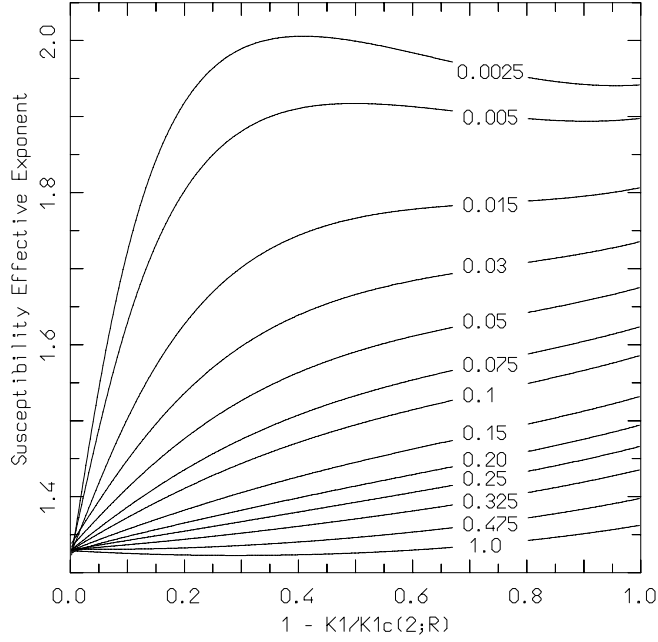}
\caption{\label{anis_f22} 
The effective exponent
$\gamma_{eff}(2;K_1,R)$, computed by PAs, for the susceptibility of
the anisotropic system is plotted vs $\tau(2;R)=1-K_1/K_{1c}(2;R)$ 
for various fixed values of  $R$
indicated on the corresponding curves.  }
\end{center}
\end{figure}

\begin{figure}[tbp]
\begin{center}
\leavevmode
\includegraphics[width=3.37 in]{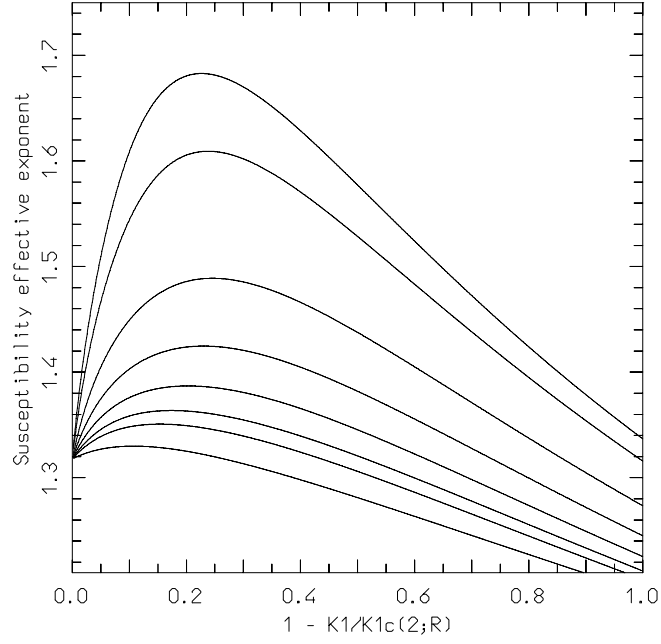}
\caption{\label{anis_feff} 
The effective exponent
$\gamma_{eff}(2;K_1,R)$ of the susceptibility,  
as obtained from the scaling form eq.(\ref{gamex}), plotted vs
$\tau(2;R)=1-K_1/K_{1c}(2;R)$.
From top to bottom $R=0.0025,0.005,0.015,0.03,0.05,0.075,0.1,0.15$.
These values of  $R$ coincide with  the eight smallest values 
 used in Fig.\ref{anis_f22}.  }
\end{center}
\end{figure}

  Let us represent the scaling function simply as $X^{(0)}(x)=P(x / \dot
 x)(1-x/ \dot x)^{-\gamma(2;R)} $, where $P(x / \dot x)$ is some function
 interpolating between the small $x$ and the large $x$ behavior of
 $X^{(0)}(x)$ and therefore taking the values $P(0)=1$ and $P(1)=\dot
 X^{(0)}$. It turns out that the simple linear expression $P(x / \dot
 x)=1+(\dot X^{(0)}-1)x/\dot x$ is a quite accurate approximation of
 the regularized scaling function $X^{(0)}(x)(1-x/\dot x)^{\gamma(2;R)} $.
 We have then used this approximate form of the crossover-scaling function
 to compute the effective exponent
\begin{equation} 
\gamma_{eff}(2;R)=\frac{K_{1c}(2;R)}{K_{1c}(2;0)}
 \frac{\tau(2;R)}{[\tau(2;0)]^{\sigma+1}}b\sigma[1+z(\frac{ P'(z)}{P(z)}+
 \frac{\gamma(2;R)}{1-z})]
\label{gamex}  
\end{equation}
with $z= x/ \dot x$. The effective exponent computed using
eq.(\ref{gamex}) is plotted vs $\tau(2;R)$ in Fig.\ref{anis_feff}, in
which we have shown the curves corresponding to the eight smallest
values of $R$ chosen in Fig.\ref{anis_f22}.  The agreement between 
  Fig.\ref{anis_f22} and Fig.\ref{anis_feff} 
  can be considered as another satisfactory confirmation of the validity
 of the crossover-scaling ansatz, if one observes that the former
 is a good PA representation of the effective
exponent on the whole interval $0<K_1<K_{1c}(2;R)$, whereas the curves
of Fig.\ref{anis_feff} can be quantitatively 
reliable only in the small range of validity
of the crossover-scaling form eq.(\ref{scalkt}), for example when
 $\tau(2;R)\lesssim 0.05$ and $R\lesssim 0.05$.

\section{Conclusions}
We have added 187 coefficients to the already known 66 coefficients of
 the bivariate HT expansion for the spin-spin correlation-function of
 the $3D$ $XY$ model, with directionally anisotropic couplings, on the
 $sc$ lattice.  Analyzing these data by PA and DA methods, we have
 determined to a good precision the critical locus of the system in
 the ferromagnetic region and checked to a fair accuracy the
 universality of the critical exponents of the susceptibility and the
 correlation-length with respect to the anisotropy parameter $R$.  We
 have also shown that the main predictions of the extended scaling
 theory for the crossover from the $3D$ to the $2D$ critical regime,
 concerning both the behavior of the line of critical points
 $K_{1c}(2;R)$ in the limit of small $R$ and the critical divergence
 of the successive $R$-derivatives, at $R=0$, of the susceptibility
 and of the second moment of the correlation-function, are compatible
 with the numerical extrapolations of our extended
 expansions. Finally, combining the HT expansions with the
 crossover-scaling ansatz, we have obtained a concrete approximate
 representation of the crossover-scaling function $X^{(0)}(x)$ for the
 susceptibility, and have shown that it reproduces, in an
 appropriately restricted temperature range, the effective exponent
 as computed by PAs on the HT side of the critical point.

\section{ APPENDIX }
For the Hamiltonian $H_{an}$ of eq.(\ref{tnhamilt}) ({\it anisotropic
 nn} model) with $z$-anisotropic  $nn$ interactions in $3D$,
 the HT expansion of the energy (and of the specific heat) 
can be simply formed in terms of the $nn$ correlation in the direction
of the  $z$-axis:

{

 \[ C(000, 0 0 1; 2 ; K_1,K_2)= K_2+
                             4 K_1^{ 2} K_2-
\frac {    1} {   2} K_2^{ 3}+
                            32 K_1^{ 4} K_2+
\frac {    1} {    3} K_2^{ 5}+
\frac {  479} {    3} K_1^{ 6} K_2+
\frac {  321} {    2} K_1^{ 4} K_2^{ 3}
                            - K_1^{ 2} K_2^{ 5}-
\frac {  11} {   48} K_2^{ 7}+  \] \[  
\frac { 1631} {    2} K_1^{ 8} K_2+
\frac { 8648} {    3} K_1^{ 6} K_2^{ 3}+
                           132 K_1^{ 4} K_2^{ 5}+
\frac {    2} {    3} K_1^{ 2} K_2^{ 7}+
\frac {   19} {  120} K_2^{ 9}+
\frac {      682193} {  180} K_1^{10} K_2+
\frac {     1278095} {   36} K_1^{ 8} K_2^{ 3}+ \] \[ 
\frac {      307798} {   27} K_1^{ 6} K_2^{ 5}-
\frac {  841} {  18} K_1^{ 4} K_2^{ 7}-
\frac {   7} {   36} K_1^{ 2} K_2^{ 9}-
\frac {  473} {4320} K_2^{11}+
\frac {      171241} {   10} K_1^{12} K_2+
\frac {     6207583} {   18} K_1^{10} K_2^{ 3}+
\frac {     3001303} {    9} K_1^{ 8} K_2^{ 5}+\] \[
\frac {      400832} {   27} K_1^{ 6} K_2^{ 7}-
\frac { 1169} {  36} K_1^{ 4} K_2^{ 9}-
\frac {    1} {  45} K_1^{ 2} K_2^{11}+
\frac {  229} { 3024} K_2^{13}+
\frac {   503352137} { 6720} K_1^{14} K_2+
\frac { 16776996529} { 5760} K_1^{12} K_2^{ 3}+\] \[
\frac {  3577027019} {  576} K_1^{10} K_2^{ 5}+
\frac {  8389362923} { 6912} K_1^{ 8} K_2^{ 7}+
\frac {     3835391} { 1728} K_1^{ 6} K_2^{ 9}+
\frac {83969} { 1920} K_1^{ 4} K_2^{11}+
\frac {  457} { 8640} K_1^{ 2} K_2^{13}-
\frac {      101369} { 1935360} K_2^{15}+
  \]  
  \[  
\frac { 77461196851} {          241920} K_1^{16} K_2+
\frac {335838080671} {15120} K_1^{14} K_2^{ 3}+
\frac {          6966586038139} {77760} K_1^{12} K_2^{ 5}+
\frac {113783648353} { 2592} K_1^{10} K_2^{ 7}+\] \[
\frac { 20055410369} {10368} K_1^{ 8} K_2^{ 9}-
\frac {    6476551} { 1440} K_1^{ 6} K_2^{11}-
\frac {     442849} {12960} K_1^{ 4} K_2^{13}-
\frac { 2483} {          90720} K_1^{ 2} K_2^{15}+
\frac {      946523} {        26127360} K_2^{17}+
  \]  
  \[  
\frac {         87581038655119} {        65318400} K_1^{18} K_2+
\frac {       2276495337319411} {        14515200} K_1^{16} K_2^{ 3}+
\frac {        656734833787141} {          604800} K_1^{14} K_2^{ 5}+
\frac {        275625926210833} {          259200} K_1^{12} K_2^{ 7}+\] \[
\frac {        196795509446797} {         1296000} K_1^{10} K_2^{ 9}+
\frac { 63925399531} {57600} K_1^{ 8} K_2^{11}-
\frac {   242287049} {  259200} K_1^{ 6} K_2^{13}+
\frac {   144905431} {         3628800} K_1^{ 4} K_2^{15}+\] \[
\frac {45679} {         7257600} K_1^{ 2} K_2^{17}-
\frac {   65467219} {      2612736000} K_2^{19}+
\frac {          3182306618881} {          576000} K_1^{20} K_2+
\frac {      34008570927861617} {        32659200} K_1^{18} K_2^{ 3}+\] \[
\frac {      84027770340130537} {         7257600} K_1^{16} K_2^{ 5}+
\frac {      13557570274734337} {          680400} K_1^{14} K_2^{ 7}+
\frac {        218753264133841} {34560} K_1^{12} K_2^{ 9}+\] \[
\frac {         39683430286691} {          144000} K_1^{10} K_2^{11}-
\frac {          114458157007} {          259200} K_1^{ 8} K_2^{13}+
\frac {  4455945869} {          907200} K_1^{ 6} K_2^{15}-
\frac {  165611879} {         2419200} K_1^{ 4} K_2^{17}+\] \[
\frac {41921} {        32659200} K_1^{ 2} K_2^{19}+
\frac {   249045899} {     14370048000} K_2^{21}+...
  \]

and of  the $nn$ correlation in the direction
of the  $x$-axis:

 \[ C(000, 1 0 0 ; 2 ; K_1,K_2)=                             K_1+
\frac {    3} {    2} K_1^{ 3}+
                             2 K_1 K_2^{ 2}+
\frac {    1} {    3} K_1^{ 5}+
                            32 K_1^{ 3} K_2^{ 2}-
\frac {  31} {   48} K_1^{ 7}+
\frac {  479} {    2} K_1^{ 5} K_2^{ 2}+
\frac {  321} {    4} K_1^{ 3} K_2^{ 4}-
\frac {    1} {   6} K_1 K_2^{ 6}+
  \]  
  \[  
\frac {  -731} { 120} K_1^{ 9}+
                          1631 K_1^{ 7} K_2^{ 2}+
                          2162 K_1^{ 5} K_2^{ 4}+
                            44 K_1^{ 3} K_2^{ 6}+
\frac {    1} {   12} K_1 K_2^{ 8}-
\frac {29239} {1440} K_1^{11}+
\frac {      682193} {   72} K_1^{ 9} K_2^{ 2}+
\frac {     1278095} {   36} K_1^{ 7} K_2^{ 4}+\]\[
\frac {      153899} {   27} K_1^{ 5} K_2^{ 6}-  
\frac {  841} {  72} K_1^{ 3} K_2^{ 8}-
\frac {    7} { 360} K_1 K_2^{10}-
\frac {     265427} { 5040} K_1^{13}+
\frac {      513723} {   10} K_1^{11} K_2^{ 2}+
\frac {    31037915} {   72} K_1^{ 9} K_2^{ 4}+\]\[
\frac {     6002606} {   27} K_1^{ 7} K_2^{ 6}+
\frac {50104} {    9} K_1^{ 5} K_2^{ 8}-
\frac { 1169} { 180} K_1^{ 3} K_2^{10}-
\frac {    1} { 540} K_1 K_2^{12}-
\frac {   75180487} {          645120} K_1^{15}+
\frac {   503352137} { 1920} K_1^{13} K_2^{ 2}+\]\[
\frac { 16776996529} { 3840} K_1^{11} K_2^{ 4}+
\frac { 17885135095} { 3456} K_1^{ 9} K_2^{ 6}+
\frac {  8389362923} {13824} K_1^{ 7} K_2^{ 8}+
\frac {     3835391} { 5760} K_1^{ 5} K_2^{10}+
\frac {83969} {11520} K_1^{ 3} K_2^{12}+\]\[
\frac {  457} {          120960} K_1 K_2^{14}-
\frac {  6506950039} {  26127360} K_1^{17}+
\frac { 77461196851} {60480} K_1^{15} K_2^{ 2}+
\frac {335838080671} { 8640} K_1^{13} K_2^{ 4}+
\frac {          6966586038139} {77760} K_1^{11} K_2^{ 6}+\]\[
\frac {568918241765} {20736} K_1^{ 9} K_2^{ 8}+
\frac { 20055410369} {25920} K_1^{ 7} K_2^{10}-
\frac {    6476551} { 5760} K_1^{ 5} K_2^{12}-
\frac {     442849} {90720} K_1^{ 3} K_2^{14}-
\frac { 2483} {  1451520} K_1 K_2^{16}+ \] \[  
\frac {          -1102473407093} {     2612736000} K_1^{19}+
\frac {         87581038655119} {        14515200} K_1^{17} K_2^{ 2}+
\frac {       2276495337319411} {         7257600} K_1^{15} K_2^{ 4}+
\frac {        656734833787141} {          518400} K_1^{13} K_2^{ 6}+\]\[
\frac {        275625926210833} {          345600} K_1^{11} K_2^{ 8}+
\frac {        196795509446797} {         2592000} K_1^{ 9} K_2^{10}+
\frac { 63925399531} {          172800} K_1^{ 7} K_2^{12}-
\frac {   242287049} {        1209600} K_1^{ 5} K_2^{14}+\]\[
\frac {   144905431} {        29030400} K_1^{ 3} K_2^{16}+
\frac {45679} {       130636800} K_1 K_2^{18}-
\frac {          6986191770643} {    14370048000} K_1^{21}+
\frac {          3182306618881} {          115200} K_1^{19} K_2^{ 2}+ \]\[  
\frac {      34008570927861617} {        14515200} K_1^{17} K_2^{ 4}+
\frac {      84027770340130537} {         5443200} K_1^{15} K_2^{ 6}+
\frac {      13557570274734337} {          777600} K_1^{13} K_2^{ 8}+\]\[ 
\frac {        218753264133841} {57600} K_1^{11} K_2^{10}+
\frac {         39683430286691} {          345600} K_1^{ 9} K_2^{12}-
\frac {114458157007} {   907200} K_1^{ 7} K_2^{14}+
\frac {  4455945869} {         4838400} K_1^{ 5} K_2^{16}+\]\[ 
\frac {   -165611879} {       21772800} K_1^{ 3} K_2^{18}+
\frac {  41921} {       653184000} K_1 K_2^{20}+...
  \]

 The HT expansion of the susceptibility $\chi(2;K_1,K_2)$ reads:

\footnotesize
 \[\chi(2;K_1,K_2)=  1+                             4 K_1+
                             2 K_2+
                            12 K_1^{ 2}+
                            16 K_1 K_2+
                             2 K_2^{ 2}+
                            34 K_1^{ 3}+
                            80 K_1^{ 2} K_2+
                            32 K_1 K_2^{ 2}+
                              K_2^{ 3}+
  \]  
  \[  
                            88 K_1^{ 4}+
                           328 K_1^{ 3} K_2+
                           240 K_1^{ 2} K_2^{ 2}+
                            40 K_1 K_2^{ 3}+
\frac {   658} {  3} K_1^{ 5}+
                          1184 K_1^{ 4} K_2+
                          1372 K_1^{ 3} K_2^{ 2}+
                           468 K_1^{ 2} K_2^{ 3}+
                            32 K_1 K_2^{ 4}+\] \[
-\frac {   1} {   3} K_2^{ 5}+  
                           529 K_1^{ 6}+
\frac {     11752} {  3} K_1^{ 5} K_2+
                          6416 K_1^{ 4} K_2^{ 2}+
                          3656 K_1^{ 3} K_2^{ 3}+
                           640 K_1^{ 2} K_2^{ 4}+
\frac { 40} {  3} K_1 K_2^{ 5}-
\frac { 1} {  6} K_2^{ 6}+
  \]  
  \[  
\frac {     14933} { 12} K_1^{ 7}+
\frac {     36484} {  3} K_1^{ 6} K_2+
                         26838 K_1^{ 5} K_2^{ 2}+
                         22161 K_1^{ 4} K_2^{ 3}+
                          6985 K_1^{ 3} K_2^{ 4}+
                           622 K_1^{ 2} K_2^{ 5}-
\frac { 8} {  3} K_1 K_2^{ 6}+\]\[
\frac {  1} { 24} K_2^{ 7}+
\frac {      5737} {  2} K_1^{ 8}+
\frac {    107813} {  3} K_1^{ 7} K_2+
                        102070 K_1^{ 6} K_2^{ 2}+
\frac {    344200} {  3} K_1^{ 5} K_2^{ 3}+
                         54480 K_1^{ 4} K_2^{ 4}+
\frac {     30092} {  3} K_1^{ 3} K_2^{ 5}+\]\[
\frac {      1202} {  3} K_1^{ 2} K_2^{ 6}-   
\frac {23} {  3} K_1 K_2^{ 7}+ 
\frac {  1} { 12} K_2^{ 8}+
\frac {    389393} { 60} K_1^{ 9}+
                        102094 K_1^{ 8} K_2+
\frac {    727969} {  2} K_1^{ 7} K_2^{ 2}+
\frac {   1585027} {  3} K_1^{ 6} K_2^{ 3}+\]\[
                        349093 K_1^{ 5} K_2^{ 4}+
\frac {    308041} {  3} K_1^{ 4} K_2^{ 5}+
\frac {     32789} {  3} K_1^{ 3} K_2^{ 6}+
\frac {629} {  6} K_1^{ 2} K_2^{ 7}-
                            4 K_1 K_2^{ 8}+
\frac {  1} { 40} K_2^{ 9}+
  \]  
  \[  
\frac {   2608499} {180} K_1^{10}+
\frac {   4211053} { 15} K_1^{ 9} K_2+
\frac {   3675469} {  3} K_1^{ 8} K_2^{ 2}+
\frac {   6689645} {  3} K_1^{ 7} K_2^{ 3}+
\frac {   3850939} {  2} K_1^{ 6} K_2^{ 4}+\]\[ 
\frac {   2446492} {  3} K_1^{ 5} K_2^{ 5}+
\frac {    913277} {  6} K_1^{ 4} K_2^{ 6}+
\frac {     26129} {  3} K_1^{ 3} K_2^{ 7}-
\frac {      293} {  3} K_1^{ 2} K_2^{ 8}+
\frac { 13} { 15} K_1 K_2^{ 9}-
\frac { 7} {360} K_2^{10}+
  \]  
  \[  
\frac {   3834323} {120} K_1^{11}+
\frac {   2253467} {  3} K_1^{10} K_2+
\frac { 355532317} { 90} K_1^{ 9} K_2^{ 2}+
\frac {  78955291} {  9} K_1^{ 8} K_2^{ 3}+
\frac {  86092934} {  9} K_1^{ 7} K_2^{ 4}+\]\[ 
\frac { 145884614} { 27} K_1^{ 6} K_2^{ 5}+
\frac {  41039243} { 27} K_1^{ 5} K_2^{ 6}+
\frac {   1607615} {  9} K_1^{ 4} K_2^{ 7}+
\frac {     38584} {  9} K_1^{ 3} K_2^{ 8}-
\frac {    11789} { 90} K_1^{ 2} K_2^{ 9}+
\frac {106} { 45} K_1 K_2^{10}+\]\[
-\frac {41} {   2160} K_2^{11}+
\frac {   1254799} { 18} K_1^{12}+
\frac { 176710657} { 90} K_1^{11} K_2+
\frac { 551185898} { 45} K_1^{10} K_2^{ 2}+
\frac { 489144733} { 15} K_1^{ 9} K_2^{ 3}+\]\[ 
\frac { 783263929} { 18} K_1^{ 8} K_2^{ 4}+
\frac { 849262729} { 27} K_1^{ 7} K_2^{ 5}+
\frac { 327546343} { 27} K_1^{ 6} K_2^{ 6}+
\frac {  62191328} { 27} K_1^{ 5} K_2^{ 7}+
\frac {   2922371} { 18} K_1^{ 4} K_2^{ 8}+\]\[ 
-\frac {      364} { 15} K_1^{ 3} K_2^{ 9}-
\frac {      793} { 15} K_1^{ 2} K_2^{10}+
\frac {283} {270} K_1 K_2^{11}-
\frac { 1} {540} K_2^{12}+
\frac {  84375807} {560} K_1^{13}+
\frac { 226326302} { 45} K_1^{12} K_2+\]\[ 
\frac { 147264673} {  4} K_1^{11} K_2^{ 2}+
\frac {6933895627} { 60} K_1^{10} K_2^{ 3}+
\frac {3332454773} { 18} K_1^{ 9} K_2^{ 4}+
\frac { 330361671} {  2} K_1^{ 8} K_2^{ 5}+\]\[
\frac {   17920524733} {216} K_1^{ 7} K_2^{ 6}+   
\frac {4843721705} {216} K_1^{ 6} K_2^{ 7}+
\frac {  17095939} {  6} K_1^{ 5} K_2^{ 8}+
\frac {   1891861} { 18} K_1^{ 4} K_2^{ 9}-
\frac {   397099} {180} K_1^{ 3} K_2^{10}+\]\[
\frac {499} { 20} K_1^{ 2} K_2^{11}-
\frac {58} {135} K_1 K_2^{12}+
\frac {191} {  30240} K_2^{13}+
\frac {6511729891} {  20160} K_1^{14}+
\frac {5315902873} {420} K_1^{13} K_2+\]\[
\frac {4839028456} { 45} K_1^{12} K_2^{ 2}+
\frac {   17700580409} { 45} K_1^{11} K_2^{ 3}+
\frac {   44549564471} { 60} K_1^{10} K_2^{ 4}+
\frac {  215942560781} {270} K_1^{ 9} K_2^{ 5}+\]\[
\frac {  108374129725} {216} K_1^{ 8} K_2^{ 6}+
\frac {   19538343311} {108} K_1^{ 7} K_2^{ 7}+
\frac {7458043925} {216} K_1^{ 6} K_2^{ 8}+
\frac { 766009373} {270} K_1^{ 5} K_2^{ 9}+\]\[
\frac {   5860253} {180} K_1^{ 4} K_2^{10}-
\frac {   261878} {135} K_1^{ 3} K_2^{11}+
\frac {      5842} {135} K_1^{ 2} K_2^{12}-
\frac {     2707} {   3780} K_1 K_2^{13}+
\frac {457} { 120960} K_2^{14}+
  \]  
  \[  
\frac {   66498259799} {  96768} K_1^{15}+
\frac {   17559467153} {560} K_1^{14} K_2+
\frac { 1030585285427} {   3360} K_1^{13} K_2^{ 2}+
\frac { 1241088903283} {960} K_1^{12} K_2^{ 3}+\]\[
\frac {  545333177917} {192} K_1^{11} K_2^{ 4}+
\frac {15628446835681} {   4320} K_1^{10} K_2^{ 5}+
\frac {11934250539301} {   4320} K_1^{ 9} K_2^{ 6}+
\frac { 4368379416899} {   3456} K_1^{ 8} K_2^{ 7}+\]\[
\frac {  127475570107} {384} K_1^{ 7} K_2^{ 8}+
\frac {  191978104529} {   4320} K_1^{ 6} K_2^{ 9}+
\frac {3117156659} {   1440} K_1^{ 5} K_2^{10}-
\frac {188490343} {   8640} K_1^{ 4} K_2^{11}+\]\[
-\frac {  4246757} {   8640} K_1^{ 3} K_2^{12}+
\frac {    114743} {   6048} K_1^{ 2} K_2^{13}-
\frac {      557} {   2520} K_1 K_2^{14}-
\frac {      109} { 193536} K_2^{15}+
  \]  
  \[  
\frac { 1054178743699} { 725760} K_1^{16}+
\frac { 1853158502071} {  24192} K_1^{15} K_2+
\frac { 5175865133419} {   6048} K_1^{14} K_2^{ 2}+
\frac {31158713948009} {   7560} K_1^{13} K_2^{ 3}+\]\[
\frac { 2498158420849} {240} K_1^{12} K_2^{ 4}+
\frac { 6679846871285} {432} K_1^{11} K_2^{ 5}+
\frac { 7582320713789} {540} K_1^{10} K_2^{ 6}+
\frac {17084567411611} {   2160} K_1^{ 9} K_2^{ 7}+\]\[
\frac { 1162221448157} {432} K_1^{ 8} K_2^{ 8}+
\frac { 2235921456439} {   4320} K_1^{ 7} K_2^{ 9}+
\frac {6389841029} {135} K_1^{ 6} K_2^{10}+
\frac {  48134516} { 45} K_1^{ 5} K_2^{11}+\]\[
-\frac { 83823949} {   2160} K_1^{ 4} K_2^{12}+
\frac {   9312019} {  15120} K_1^{ 3} K_2^{13}-
\frac {   233021} {  30240} K_1^{ 2} K_2^{14}+
\frac {      3257} {  17280} K_1 K_2^{15}-
\frac {     2483} {1451520} K_2^{16}+\]\[
\frac {39863505993331} {13063680} K_1^{17}+
\frac { 4789948744531} {  25920} K_1^{16} K_2+
\frac {566876641827121} { 241920} K_1^{15} K_2^{ 2}+
\frac {773547561026857} {  60480} K_1^{14} K_2^{ 3}+\]\[
\frac {21204512261149} {576} K_1^{13} K_2^{ 4}+
\frac {   4893241012078331} {  77760} K_1^{12} K_2^{ 5}+
\frac {   2601671974782203} {  38880} K_1^{11} K_2^{ 6}+\]\[
\frac {585636405603727} {  12960} K_1^{10} K_2^{ 7}+
\frac {994750117740587} {  51840} K_1^{ 9} K_2^{ 8}+
\frac {254490705579917} {  51840} K_1^{ 8} K_2^{ 9}+\]\[
\frac {17805962398507} {  25920} K_1^{ 7} K_2^{10}+
\frac {  261786374297} {   6480} K_1^{ 6} K_2^{11}+
\frac { 568937123} {  25920} K_1^{ 5} K_2^{12}-
\frac {877704563} {  36288} K_1^{ 4} K_2^{13}+\]\[
\frac {  27164849} {  36288} K_1^{ 3} K_2^{14}-
\frac { 11452451} { 725760} K_1^{ 2} K_2^{15}+
\frac {     18229} {  90720} K_1 K_2^{16}-
\frac {    16151} {26127360} K_2^{17}+
\frac {19830277603399} {3110400} K_1^{18}+\]\[
\frac {   1439510708686567} {3265920} K_1^{17} K_2+
\frac {   2287661692966927} { 362880} K_1^{16} K_2^{ 2}+
\frac {   4687330747502107} { 120960} K_1^{15} K_2^{ 3}+\]\[
\frac {   3810140897792597} {  30240} K_1^{14} K_2^{ 4}+
\frac {  66815169002499599} { 272160} K_1^{13} K_2^{ 5}+
\frac {  46842552495817169} { 155520} K_1^{12} K_2^{ 6}+\]\[
\frac {   4658208957065401} {  19440} K_1^{11} K_2^{ 7}+
\frac {   6388084191664099} {  51840} K_1^{10} K_2^{ 8}+
\frac { 6411755638079} {160} K_1^{ 9} K_2^{ 9}+\]\[
\frac {133534055859457} {  17280} K_1^{ 8} K_2^{10}+
\frac { 1997340559319} {   2592} K_1^{ 7} K_2^{11}+
\frac {  261618690883} {  10368} K_1^{ 6} K_2^{12}-
\frac {  10212267019} {  18144} K_1^{ 5} K_2^{13}+\]\[
-\frac { 49463021} {  90720} K_1^{ 4} K_2^{14}+
\frac {  30377351} { 120960} K_1^{ 3} K_2^{15}-
\frac {   111247} {  17280} K_1^{ 2} K_2^{16}+
\frac {     95791} {3265920} K_1 K_2^{17}+
\frac {     45679} {   130636800} K_2^{18}+\]\[
\frac {   8656980509809027} {   653184000} K_1^{19}+    
\frac {  16990346289372257} {16329600} K_1^{18} K_2+
\frac { 181822163330740067} {10886400} K_1^{17} K_2^{ 2}+\]\[
\frac { 277976926650925547} {2419200} K_1^{16} K_2^{ 3}+
\frac { 253619834611062257} { 604800} K_1^{15} K_2^{ 4}+
\frac { 139504244168782903} { 151200} K_1^{14} K_2^{ 5}+\]\[
\frac { 586373897113473881} { 453600} K_1^{13} K_2^{ 6}+
\frac { 154458497864612623} { 129600} K_1^{12} K_2^{ 7}+
\frac { 188558667394136629} { 259200} K_1^{11} K_2^{ 8}+\]\[
\frac { 188013182402455507} { 648000} K_1^{10} K_2^{ 9}+
\frac {982091938311529} {  13500} K_1^{ 9} K_2^{10}+
\frac {908320442109053} {  86400} K_1^{ 8} K_2^{11}+\]\[
\frac {30871722528221} {  43200} K_1^{ 7} K_2^{12}+
\frac { 6909844908217} { 907200} K_1^{ 6} K_2^{13}-
\frac { 131184772823} { 226800} K_1^{ 5} K_2^{14}+
\frac {3786229351} { 302400} K_1^{ 4} K_2^{15}+\]\[
-\frac {167790391} { 806400} K_1^{ 3} K_2^{16}+
\frac {  18884699} {3628800} K_1^{ 2} K_2^{17}-
\frac {81803} {1166400} K_1 K_2^{18}+
\frac {529301} {1306368000 } K_2^{19}+\]\[
\frac {2985467351081077} {108864000} K_1^{20}+
\frac {  397290645875080747} {163296000} K_1^{19} K_2+
\frac {   31654153295670463} {   725760} K_1^{18} K_2^{ 2}+\]\[
\frac { 1091318923155197137} {  3265920} K_1^{17} K_2^{ 3}+
\frac {  234970051697118173} {   172800} K_1^{16} K_2^{ 4}+
\frac {18262470593869394021} {  5443200} K_1^{15} K_2^{ 5}+\]\[
\frac {14461460147180084629} {  2721600} K_1^{14} K_2^{ 6}+
\frac { 5092048288872136451} {   907200} K_1^{13} K_2^{ 7}+
\frac { 3104846020933512733} {   777600} K_1^{12} K_2^{ 8}+\]\[
\frac { 1854733050821010329} {   972000} K_1^{11} K_2^{ 9}+
\frac {   96563772320850199} {   162000} K_1^{10} K_2^{10}+
\frac {   37397800587256709} {   324000} K_1^{ 9} K_2^{11}+\]\[
\frac {3185944243721879} {   259200} K_1^{ 8} K_2^{12}+
\frac {   2088234805193} {     4032} K_1^{ 7} K_2^{13}-
\frac {  1262744580547} {   226800} K_1^{ 6} K_2^{14}+\]\[
-\frac {   222365630579} {   907200} K_1^{ 5} K_2^{15}+
\frac {     40181717827} {  3628800} K_1^{ 4} K_2^{16}-
\frac {     4507243337} { 16329600} K_1^{ 3} K_2^{17}+\]\[
\frac {   174167243} { 32659200} K_1^{ 2} K_2^{18}
-\frac {    8229559} {163296000} K_1 K_2^{19}+
\frac {   41921} {653184000} K_2^{20}+\]
  \[  
\frac {   811927408684296587} {   14370048000} K_1^{21}+
\frac {     4385492759351981} { 777600} K_1^{20} K_2^+
\frac { 12241208160951859717} {     108864000} K_1^{19} K_2^{ 2}+\]\[
\frac { 62360406854392578737} {      65318400} K_1^{18} K_2^{ 3}+
\frac {  3128332273134190457} { 725760} K_1^{17} K_2^{ 4}+
\frac { 17192353387793409331} {1451520} K_1^{16} K_2^{ 5}+\]\[
\frac {  4073641693391130869} { 193536} K_1^{15} K_2^{ 6}+
\frac {548535341435511268981} {      21772800} K_1^{14} K_2^{ 7}+
\frac {149677215569420577263} {7257600} K_1^{13} K_2^{ 8}+\]\[
\frac { 22488614102184820471} {1944000} K_1^{12} K_2^{ 9}+
\frac {     7882602902572789} {1800} K_1^{11} K_2^{10}+
\frac {   698300238344288257} { 648000} K_1^{10} K_2^{11}+\]\[
\frac {      721052835977723} {4500} K_1^{ 9} K_2^{12}+
\frac {    88211417051098193} {7257600} K_1^{ 8} K_2^{13}+
\frac {      253014314268331} {1036800} K_1^{ 7} K_2^{14}+\]\[
\frac {-5964052451177} { 580608} K_1^{ 6} K_2^{15}+
\frac {85859357495} { 870912} K_1^{ 5} K_2^{16}+
\frac {11197132927} {   4665600} K_1^{ 4} K_2^{17}+\]\[
\frac {-5286752777} {      65318400} K_1^{ 3} K_2^{18}-
\frac {  39273503} {      46656000} K_1^{ 2} K_2^{19}+
\frac {      36697} {      13608000} K_1 K_2^{20}-
\frac {    168103} {    1368576000} K_2^{21}+...
  \]

The expansion of the second moment of the correlation-function
 $m^{(2)}(2;K_1,K_2)$ reads:

 \[ m^{(2)}(2;K_1,K_2)=                               4 K_1+
                             2 K_2+
                            32 K_1^{ 2}+
                            32 K_1 K_2+
                             8 K_2^{ 2}+
                           162 K_1^{ 3}+
                           272 K_1^{ 2} K_2+
                           128 K_1 K_2^{ 2}+
                            17 K_2^{ 3}+
  \]  
  \[  
                           672 K_1^{ 4}+
                          1680 K_1^{ 3} K_2+
                          1248 K_1^{ 2} K_2^{ 2}+
                           336 K_1 K_2^{ 3}+
                            24 K_2^{ 4}+
\frac {  7378} {  3} K_1^{ 5}+ 
                          8544 K_1^{ 4} K_2+\] \[
                          9100 K_1^{ 3} K_2^{ 2}+
                          3892 K_1^{ 2} K_2^{ 3}+
                           640 K_1 K_2^{ 4}+
\frac {    71} {  3} K_2^{ 5}+
\frac { 24772} {  3} K_1^{ 6}+ 
\frac {114128} {  3} K_1^{ 5} K_2+
                         54848 K_1^{ 4} K_2^{ 2}+\] \[
                         33552 K_1^{ 3} K_2^{ 3}+
                          9088 K_1^{ 2} K_2^{ 4}+
\frac {  2768} {  3} K_1 K_2^{ 5}+
\frac {    46} {  3} K_2^{ 6}+
\frac {312149} { 12} K_1^{ 7}+
\frac {460804} {  3} K_1^{ 6} K_2+\] \[
                        287078 K_1^{ 5} K_2^{ 2}+
                        236593 K_1^{ 4} K_2^{ 3}+
                         93241 K_1^{ 3} K_2^{ 4}+
                         16622 K_1^{ 2} K_2^{ 5}+
\frac {  3040} {  3} K_1 K_2^{ 6}+
\frac {    97} { 24} K_2^{ 7}+
                         77996 K_1^{ 8}+\] \[
\frac {   1725322} {  3} K_1^{ 7} K_2+
\frac {   4053232} {  3} K_1^{ 6} K_2^{ 2}+
                       1437104 K_1^{ 5} K_2^{ 3}+
                        768088 K_1^{ 4} K_2^{ 4}+
\frac {616568} {  3} K_1^{ 3} K_2^{ 5}+\] \[
\frac { 72896} {  3} K_1^{ 2} K_2^{ 6}+
\frac {  2450} {  3} K_1 K_2^{ 7} -
\frac {   11} {  3} K_2^{ 8}+
\frac {  13484753} { 60} K_1^{ 9}+
\frac {   6082178} {  3} K_1^{ 8} K_2+\] \[
\frac {  11686645} {  2} K_1^{ 7} K_2^{ 2}+
\frac {  23325767} {  3} K_1^{ 6} K_2^{ 3}+
                       5379749 K_1^{ 5} K_2^{ 4}+
\frac {   5953681} {  3} K_1^{ 4} K_2^{ 5}+\] \[
                        369267 K_1^{ 3} K_2^{ 6}+
\frac {170477} {  6} K_1^{ 2} K_2^{ 7}+
\frac {  1232} {  3} K_1 K_2^{ 8}-
\frac {  199} { 40} K_2^{ 9}+  
\frac {  28201211} { 45} K_1^{10}+\] \[
\frac {  34012382} {  5} K_1^{ 9} K_2+
\frac {  70813492} {  3} K_1^{ 8} K_2^{ 2}+
                      38343886 K_1^{ 7} K_2^{ 3}+
                      33223346 K_1^{ 6} K_2^{ 4}+\] \[
\frac {  48053336} {  3} K_1^{ 5} K_2^{ 5}+
                       4215454 K_1^{ 4} K_2^{ 6}+
\frac {   1642850} {  3} K_1^{ 3} K_2^{ 7}+
\frac { 77800} {  3} K_1^{ 2} K_2^{ 8}+
\frac {   106} { 15} K_1 K_2^{ 9}+\] \[
-\frac {   187} {90} K_2^{10}+
\frac { 611969977} {360} K_1^{11}+
\frac { 985452181} { 45} K_1^{10} K_2+
\frac {1621421069} { 18} K_1^{ 9} K_2^{ 2}+\] \[
\frac {1577314567} {  9} K_1^{ 8} K_2^{ 3}+
\frac {1668258680} {  9} K_1^{ 7} K_2^{ 4}+
\frac {3039023000} { 27} K_1^{ 6} K_2^{ 5}+
\frac {1061738147} { 27} K_1^{ 5} K_2^{ 6}+\] \[
\frac {  67469699} {  9} K_1^{ 4} K_2^{ 7}+
\frac {   6031258} {  9} K_1^{ 3} K_2^{ 8}+
\frac {   1531267} { 90} K_1^{ 2} K_2^{ 9}-
\frac {  9176} {45} K_1 K_2^{10}+\] \[
\frac {   395} {432} K_2^{11}+
\frac { 202640986} { 45} K_1^{12}+
\frac { 612509213} {  9} K_1^{11} K_2+
\frac {   14744551642} { 45} K_1^{10} K_2^{ 2}+\] \[
\frac {   11275330394} { 15} K_1^{ 9} K_2^{ 3}+
\frac {8562260260} {  9} K_1^{ 8} K_2^{ 4}+
\frac {   19108224278} { 27} K_1^{ 7} K_2^{ 5}+
\frac {2830398088} {  9} K_1^{ 6} K_2^{ 6}+\] \[
\frac {2196959680} { 27} K_1^{ 5} K_2^{ 7}+
\frac { 101314112} {  9} K_1^{ 4} K_2^{ 8}+
\frac {   3327988} {  5} K_1^{ 3} K_2^{ 9}+
\frac { 82946} { 15} K_1^{ 2} K_2^{10}+\] \[
-\frac { 25061} { 135} K_1 K_2^{11}+
\frac {   457} {270} K_2^{12}+
\frac {   58900571047} {   5040} K_1^{13}+
\frac {9232198298} { 45} K_1^{12} K_2+\] \[
\frac {  205863030521} {180} K_1^{11} K_2^{ 2}+
\frac {  549781840789} {180} K_1^{10} K_2^{ 3}+
\frac {   81925739897} { 18} K_1^{ 9} K_2^{ 4}+
\frac {   73086198751} { 18} K_1^{ 8} K_2^{ 5}+\] \[
\frac {  480629562289} {216} K_1^{ 7} K_2^{ 6}+
\frac {  160430759525} {216} K_1^{ 6} K_2^{ 7}+
\frac {2587900213} { 18} K_1^{ 5} K_2^{ 8}+
\frac { 256823701} { 18} K_1^{ 4} K_2^{ 9}+\] \[
\frac {  91921193} {180} K_1^{ 3} K_2^{10}-
\frac {39139} { 12} K_1^{ 2} K_2^{11}-
\frac { 6856} {135} K_1 K_2^{12}+
\frac { 22199} {  30240} K_2^{13}+
  \]  
  \[  
\frac {3336209179} {112} K_1^{14}+
\frac {   42155577203} { 70} K_1^{13} K_2+
\frac {   19258430024} {  5} K_1^{12} K_2^{ 2}+
\frac {  177682250558} { 15} K_1^{11} K_2^{ 3}+\] \[
\frac {  307643703989} { 15} K_1^{10} K_2^{ 4}+
\frac { 2911760805313} {135} K_1^{ 9} K_2^{ 5}+
\frac {  768911042053} { 54} K_1^{ 8} K_2^{ 6}+
\frac {  319585153487} { 54} K_1^{ 7} K_2^{ 7}+\] \[
\frac {   27117476941} { 18} K_1^{ 6} K_2^{ 8}+
\frac {9822059771} { 45} K_1^{ 5} K_2^{ 9}+
\frac { 135760609} {  9} K_1^{ 4} K_2^{10}+
\frac {   6957800} { 27} K_1^{ 3} K_2^{11}+\] \[
-\frac {   837212} {135} K_1^{ 2} K_2^{12}+
\frac {106661} {   1890} K_1 K_2^{13}-
\frac { 9959} {  30240} K_2^{14}+
\frac { 1721567587879} {  23040} K_1^{15}+
\frac { 1740519488525} {   1008} K_1^{14} K_2+\] \[
\frac { 126704980674473} {  10080} K_1^{13} K_2^{ 2}+
\frac {42345007132019} {960} K_1^{12} K_2^{ 3}+
\frac {84243175256177} {960} K_1^{11} K_2^{ 4}+\] \[
\frac { 463449944770897} {   4320} K_1^{10} K_2^{ 5}+
\frac { 361807221213917} {   4320} K_1^{ 9} K_2^{ 6}+
\frac { 145911608744899} {   3456} K_1^{ 8} K_2^{ 7}+
\frac {46852875522355} {   3456} K_1^{ 7} K_2^{ 8}+\] \[
\frac {11447940964657} {   4320} K_1^{ 6} K_2^{ 9}+
\frac {  409797896227} {   1440} K_1^{ 5} K_2^{10}+
\frac {  111235514009} {   8640} K_1^{ 4} K_2^{11}+
\frac {  35705321} {   2880} K_1^{ 3} K_2^{12}+\] \[
-\frac { 122203853} { 30240} K_1^{ 2} K_2^{13}+
\frac {134969} {   1890} K_1 K_2^{14}-
\frac {   185291} { 322560} K_2^{15}+
\frac {16763079262169} {  90720} K_1^{16}+
\frac { 293228729314067} {  60480} K_1^{15} K_2+\] \[
\frac {75400838218273} {   1890} K_1^{14} K_2^{ 2}+
\frac { 599027907233393} {   3780} K_1^{13} K_2^{ 3}+
\frac {14354418431047} { 40} K_1^{12} K_2^{ 4}+\] \[
\frac { 544406144647673} {   1080} K_1^{11} K_2^{ 5}+
\frac {49513057580797} {108} K_1^{10} K_2^{ 6}+
\frac { 6591403549631} { 24} K_1^{ 9} K_2^{ 7}+
\frac {46629805507009} {432} K_1^{ 8} K_2^{ 8}+\] \[
\frac {58466928681791} {   2160} K_1^{ 7} K_2^{ 9}+
\frac {  730982464631} {180} K_1^{ 6} K_2^{10}+
\frac {   42641216914} {135} K_1^{ 5} K_2^{11}+
\frac { 973964951} {120} K_1^{ 4} K_2^{12}+\] \[
-\frac { 139288171} { 1080} K_1^{ 3} K_2^{13}-
\frac {  451397} {   1890} K_1^{ 2} K_2^{14}+
\frac {508229} {  20160} K_1 K_2^{15}-
\frac {15487} {  72576} K_2^{16}+
  \]  
  \[  
\frac {5893118865913171} { 13063680} K_1^{17}+
\frac {2424120194567633} { 181440} K_1^{16} K_2+
\frac {   29877830035592213} { 241920} K_1^{15} K_2^{ 2}+\] \[
\frac {   33355738800284053} {  60480} K_1^{14} K_2^{ 3}+
\frac {   85279687665510997} {  60480} K_1^{13} K_2^{ 4}+
\frac {  175231349967878723} {  77760} K_1^{12} K_2^{ 5}+\] \[
\frac {   91688471004674387} {  38880} K_1^{11} K_2^{ 6}+
\frac {   21406776018978923} {  12960} K_1^{10} K_2^{ 7}+
\frac {8051864567496943} {  10368} K_1^{ 9} K_2^{ 8}+\] \[
\frac {   12502971065814437} {  51840} K_1^{ 8} K_2^{ 9}+
\frac { 410022854780933} {   8640} K_1^{ 7} K_2^{10}+
\frac {11705929429057} {   2160} K_1^{ 6} K_2^{11}+\] \[
\frac { 7560435703027} {  25920} K_1^{ 5} K_2^{12}+
\frac {  465465383273} { 181440} K_1^{ 4} K_2^{13}-
\frac {   4996914619} {  36288} K_1^{ 3} K_2^{14}+\] \[
\frac {1454610517} { 725760} K_1^{ 2} K_2^{15}-
\frac {   415019} {  22680} K_1 K_2^{16}+
\frac {724301} {5225472} K_2^{17}+
\frac {   17775777329026559} { 16329600} K_1^{18}+\] \[
\frac {   59110336698456487} {1632960} K_1^{17} K_2+
\frac {   33916755588333461} {  90720} K_1^{16} K_2^{ 2}+
\frac {  112823528586303187} {  60480} K_1^{15} K_2^{ 3}+\] \[
\frac {  121270297167625891} {  22680} K_1^{14} K_2^{ 4}+
\frac {  262383267925844563} {  27216} K_1^{13} K_2^{ 5}+
\frac {  446682230590568881} {  38880} K_1^{12} K_2^{ 6}+\] \[
\frac {   18047652078347753} {   1944} K_1^{11} K_2^{ 7}+
\frac {   66470311691332679} {  12960} K_1^{10} K_2^{ 8}+
\frac { 461244600701317} {240} K_1^{ 9} K_2^{ 9}+\] \[
\frac {2053612405901317} {   4320} K_1^{ 8} K_2^{10}+
\frac {95019500270231} {   1296} K_1^{ 7} K_2^{11}+
\frac { 1801642236389} {288} K_1^{ 6} K_2^{12}+\] \[
\frac { 1068304780441} {   5040} K_1^{ 5} K_2^{13}-
\frac {356581981} {216} K_1^{ 4} K_2^{14}-
\frac { 3715798969} {  60480} K_1^{ 3} K_2^{15}+
\frac {  27612313} {  15120} K_1^{ 2} K_2^{16}+\] \[
-\frac { 42818873} {1632960} K_1 K_2^{17}+
\frac {   6111499} { 32659200} K_2^{18}+
\frac { 1697692411053976387} {653184000} K_1^{19}+
\frac {  225292842432658127} {2332800} K_1^{18} K_2+\] \[
\frac {12073089792078645727} { 10886400} K_1^{17} K_2^{ 2}+
\frac {14877683079746729947} {2419200} K_1^{16} K_2^{ 3}+
\frac {35635183308116238361} {1814400} K_1^{15} K_2^{ 4}+\] \[
\frac {17990028013859879689} { 453600} K_1^{14} K_2^{ 5}+
\frac {24197213631925141831} { 453600} K_1^{13} K_2^{ 6}+
\frac { 6374234328455740363} { 129600} K_1^{12} K_2^{ 7}+\] \[
\frac {  906251536245123461} {  28800} K_1^{11} K_2^{ 8}+
\frac { 9035338464848376767} { 648000} K_1^{10} K_2^{ 9}+
\frac {   85175951228570221} {  20250} K_1^{ 9} K_2^{10}+\] \[
\frac {   23941624054521991} {  28800} K_1^{ 8} K_2^{11}+
\frac {4310281756047181} {  43200} K_1^{ 7} K_2^{12}+
\frac {5571935169950327} { 907200} K_1^{ 6} K_2^{13}+\] \[
\frac {23050830711047} { 226800} K_1^{ 5} K_2^{14}-
\frac { 419715210991} { 129600} K_1^{ 4} K_2^{15}+
\frac {   48764703667} {2419200} K_1^{ 3} K_2^{16}+\] \[
\frac {4501185937} { 10886400} K_1^{ 2} K_2^{17}-
\frac { 18779261} {2041200} K_1 K_2^{18}+
\frac {  64742261} {   1306368000 } K_2^{19}+
\frac {  41816028466101527} { 6804000} K_1^{20}+\] \[
\frac {      20744074405363158427} {81648000} K_1^{19} K_2+
\frac {3296051006124220457} { 1020600} K_1^{18} K_2^{ 2}+
\frac { 399315806473425521} {   20160} K_1^{17} K_2^{ 3}+\] \[
\frac {9084520466090723077} {  129600} K_1^{16} K_2^{ 4}+
\frac {     428717807069556189091} { 2721600} K_1^{15} K_2^{ 5}+
\frac {     161475419175188528617} {  680400} K_1^{14} K_2^{ 6}+\] \[
\frac { 593574184668411599} {    2400} K_1^{13} K_2^{ 7}+
\frac {      17599421384527454017} {   97200} K_1^{12} K_2^{ 8}+
\frac {5041937738484379241} {   54000} K_1^{11} K_2^{ 9}+\] \[
\frac {5438875343509708877} {  162000} K_1^{10} K_2^{10}+
\frac {1328632848231694729} {  162000} K_1^{ 9} K_2^{11}+
\frac {  20953128568172143} {   16200} K_1^{ 8} K_2^{12}+\] \[
\frac {  10792829938226569} {   90720} K_1^{ 7} K_2^{13}+
\frac {   1119984288812191} {  226800} K_1^{ 6} K_2^{14}+
\frac {   187714666081} {  453600} K_1^{ 5} K_2^{15}+\] \[
-\frac { 1090393669307} {  453600} K_1^{ 4} K_2^{16}+
\frac {   434088485513} { 8164800} K_1^{ 3} K_2^{17}-
\frac {    2753717513} { 4082400} K_1^{ 2} K_2^{18}+\] \[
\frac {562928201} {81648000} K_1 K_2^{19}-
\frac {    9330469} {    163296000} K_2^{20}+
\frac {206973837048951639371} {   14370048000} K_1^{21}+\] \[
\frac { 53873357653341570569} {  81648000} K_1^{20} K_2+
\frac { 40267035928596854137} {  4354560} K_1^{19} K_2^{ 2}+
\frac {  4079484985047403119493} {  65318400} K_1^{18} K_2^{ 3}+\] \[
\frac {379052188965132965269} { 1555200} K_1^{17} K_2^{ 4}+
\frac { 13196982051156523866581} {21772800} K_1^{16} K_2^{ 5}+
\frac { 44245435623157935554441} {43545600} K_1^{15} K_2^{ 6}+\] \[
\frac {  8616554448586239218387} { 7257600} K_1^{14} K_2^{ 7}+
\frac { 21437939380897068658541} {21772800} K_1^{13} K_2^{ 8}+
\frac {  1133763862909311000887} { 1944000} K_1^{12} K_2^{ 9}+\] \[
\frac { 39776679878739884231} {  162000} K_1^{11} K_2^{10}+
\frac { 15594884456933809459} {  216000} K_1^{10} K_2^{11}+
\frac {    15458199431760647} {1080} K_1^{ 9} K_2^{12}+\] \[
\frac {   173186300601405355} {   96768} K_1^{ 8} K_2^{13}+
\frac {   890557841004984769} { 7257600} K_1^{ 7} K_2^{14}+
\frac {    42723412499535059} {14515200} K_1^{ 6} K_2^{15}+\] \[
\frac {  -1259434901063897} { 21772800} K_1^{ 5} K_2^{16}-
\frac {17420157160631} {32659200} K_1^{ 4} K_2^{17}+
\frac {   774855209473} {21772800} K_1^{ 3} K_2^{18}+\] \[
\frac {-35002695431} { 46656000} K_1^{ 2} K_2^{19}+
\frac {   13528453} {  1458000} K_1 K_2^{20}-
\frac { 108949189} {    1916006400} K_2^{21}+...
 \]
    
}

\section{Acknowledgements}
 We thank  Prof. M. E. Fisher for his interest and
 encouragement.  We have enjoyed the hospitality and support 
of the Physics Depts. of Milano-Bicocca University and of Milano University. 
 Our calculations were partly performed  by the {\it Turing}
 {\it pc}-cluster of the Sezione INFN of Milano-Bicocca.  
Partial support by the MIUR is  acknowledged.

\newpage


\footnotesize

\begin{table}
\caption{
Estimates of the critical inverse-temperatures of the 
anisotropic system  for various values of $R$. 
Only the two smallest-$R$ estimates are biased, the remaining ones
 being  unbiased.} 
 \center
\begin{tabular}{|c|c|c|c|c|c|}
 \hline
$R$ &  $K_{1c}(2;R) $ &$R$ &  $K_{1c}(2;R) $&$R$ &  $K_{1c}(2;R) $   \\
 \hline
3.4&0.13901(6)   &1.3 &0.20727(3)&0.15 & 0.3562(3) \\
3.0 &0.14724(6) &1.0&0.22710(3) &  0.1 & 0.3776(5)\\
2.6 &0.15693(6) & 0.8&0.24394(3) &  0.075 &0.391(2)\\
2.2 &0.16861(5)  &0.6 &0.26537(3) &  0.05 &0.409(3) \\
1.9 &0.17914(5)  &0.4 &0.29460(6) & 0.0125&0.454(4)  \\
1.6 &0.19176(4)  &0.2&0.3395(2)& 0.005&0.473(4)   \\
\colrule  
\end{tabular} 
\label{tabella}
\end{table}

\begin{table}
\caption{
Estimates, by first-order DAs, of the critical values  $G^{(0)}_s $ 
of the normalized ratios $G^{(0)}_s(K_1) $ as  $K_1 \rightarrow K_{1c}(2;0)$.} 
 \center
\begin{tabular}{|c|c|c|c|c|c|c|}
 \hline
critical value&$G^{(0)}_1 $&$G^{(0)}_2$& $G^{(0)}_3$& $G^{(0)}_4$&$G^{(0)}_5$
&$G^{(0)}_6$\\
 \hline
&1.& 1.00(1) &1.01(1)  &1.00(2) &1.02(8)&0.95(15) \\
 \hline
critical value &$G^{(2)}_1$&$G^{(2)}_2$&$G^{(2)}_3 $&$G^{(2)}_4$&$G^{(2)}_5$ 
&$G^{(2)}_6$\\
 \hline
&1.04(5)& .99(4) &0.99(1)  &0.99(1) & .99(3)&0.98(13) \\

\colrule  
\end{tabular} 
\label{tabella2}
\end{table}

\begin{table}
\caption{
 Estimates of the universal ratios $Q_s$, computed by extrapolating 
first-order DAs of the effective ratios in eq.(\ref{effrat}).  
 } 
 \center
\begin{tabular}{|c|c|c|c|c|c|}
 \hline
$Q_1$&$Q_2$&$Q_3$&$Q_4$&$Q_5$&$Q_6$\\
 \hline
 1.584(6) &1.436(2)&1.287(9) &1.24(1) &1.18(1)&1.12(9)\\
 \hline
\colrule  
\end{tabular} 
\label{tabella4}
\end{table}

\end{document}